\documentclass[journal]{IEEEtran}

\ifCLASSINFOpdf
\else
\fi
\PassOptionsToPackage{table, dvipsnames}{xcolor}
\usepackage[linesnumbered,ruled,vlined]{algorithm2e}

\newlength\mylen

\usepackage{graphicx}
\usepackage{float}
\usepackage{multirow}
\usepackage{hhline}
\usepackage{colortbl}
\usepackage{framed}
\usepackage{mdframed}
\usepackage{amsmath}
\usepackage{array}
\usepackage[hyphens]{url}
\usepackage{hyperref}
\usepackage{subcaption}
\newcommand{\ccomment}[1]{\textcolor{OliveGreen}{#1}}
\usepackage[table,xcdraw]{xcolor}
\usepackage{listings}
\definecolor{darkred}{rgb}{0.6,0.0,0.0}
\definecolor{mycolor1}{RGB}{0, 0, 180}
\definecolor{mycolor2}{RGB}{31, 132, 31}
\lstdefinestyle{customkeywords}{
    language=Python,
    basicstyle=\ttfamily\small,
    keywordstyle=\color{mycolor1},  
    morekeywords={Environment, Road_network, Participant, Criteria},  
    moredelim=**[is][\color{red}]{@@}{@@},
    moredelim=**[is][\color{mycolor2}]{@}{@},  
}

\usepackage{tabularx}

\newcommand{\tool}{\textsc{TARGET}}

\begin{document}

\title{TARGET: Traffic Rule-based Test Generation for Autonomous Driving via Validated LLM-Guided Knowledge Extraction}

\author{Yao Deng, Zhi Tu, Jiaohong Yao,  Mengshi Zhang, Tianyi Zhang, Xi Zheng

\IEEEcompsocitemizethanks{\IEEEcompsocthanksitem Y. Deng, J. Yao, and  X. Zheng are with the School
of Computing, Macquarie University, Sydney,
NSW. \protect
E-mail: yao.deng@hdr.mq.edu.au, jiaohong.yao@hdr.mq.edu.au, james.zheng@mq.edu.au; Xi Zheng is the corresponding author
\IEEEcompsocthanksitem Z, Tu, and T. Zhang are with the Department of Computer Science, Purdue University, West Lafayette, IN. \protect
E-mail: zhitu@purdue.edu, tianyi@purdue.edu
\IEEEcompsocthanksitem M. Zhang is with TensorBlock, 254 Chapman Rd, Newark, DE. \protect
E-mail: mengshiz@tensorblock.co
}
}

\maketitle
\begin{abstract}
Recent incidents with autonomous vehicles highlight the need for rigorous testing to ensure safety and robustness. Constructing test scenarios for autonomous driving systems (ADSs), however, is labor-intensive. We propose {\tool}, an end-to-end framework that automatically generates test scenarios from traffic rules. To address complexity, we leverage a Large Language Model (LLM) to extract knowledge from traffic rules. To mitigate hallucinations caused by large context during input processing, we introduce a domain-specific language (DSL) designed to be syntactically simple and compositional. This design allows the LLM to learn and generate test scenarios in a modular manner while enabling syntactic and semantic validation for each component. Based on these validated representations, {\tool} synthesizes executable scripts to render scenarios in simulation. Evaluated seven ADSs with 284 scenarios derived from 54 traffic rules, {\tool} uncovered 610 rule violations, collisions, and other issues. For each violation, {\tool} generates scenario recordings and detailed logs, aiding root cause analysis. Two identified issues were confirmed by ADS developers: one linked to an existing bug report and the other to limited ADS functionality.
\end{abstract}

\begin{IEEEkeywords}
Test Generation, Autonomous Driving System Testing, Scenario-based Testing, LLM
\end{IEEEkeywords}

\section{Introduction}
\label{sec:introduction}

Systematic testing is vital for ensuring the robustness and safety of autonomous driving systems (ADSs) across diverse scenarios. In some regions, autonomous vehicles must pass assessments involving hundreds of scenarios derived from traffic rules and human driving experiences~\cite{vehicle_test}. To evaluate ADSs before on-road testing, simulation testing has become prevalent~\cite{lou2021testing, huang2016autonomous, ma2020artificial, birchler2024roadmap}. Simulators like CARLA~\cite{dosovitskiy2017carla} and LGSVL~\cite{rong2020lgsvl} provide safe, customizable environments for generating test scenarios~\cite{sun2022lawbreaker, zhou2023specification, li2020av}. In recent research~\cite{birchler2024roadmap} about the roadmap of simulation-based ADS testing, how to define test scenarios in the simulation environment is identified as a key challenge.
In current industry practices, testers manually define test scenarios and translate them into simulator-supported configurations using complex Domain-Specific Languages (DSLs) like OpenSCENARIO~\cite{openScenario}, which require detailed parameter specifications and result in verbose, error-prone representations. For instance, describing a lane-change scenario in OpenSCENARIO involves approximately 260 lines of code~\footnote{\href{https://github.com/carla-simulator/scenario_runner/blob/master/srunner/examples/LaneChangeSimple.xosc}{Lane-changing scenario example in OpenSCENARIO DSL syntax\label{fn_dsl}}.}. This highlights the need for automated frameworks to simplify scenario representation and generate test scenarios from resources like traffic rules~\cite{lou2021testing}. 

Existing efforts include RMT~\cite{deng2020rmt}, which uses traditional NLP techniques to extract information from simplified traffic rules rewritten by testers, but these fail to capture real-world complexity and struggle with deep understanding. LawBreaker~\cite{sun2022lawbreaker} encodes traffic rules with Signal Temporal Logic (STL) and applies fuzzing to generate violating scenarios, yet manual translation into STL is time-intensive and complex. While Large Language Models (LLMs) have shown promise, the intricate grammar and lack of compositional structure in DSLs like OpenSCENARIO and Scenic make extracting knowledge with LLMs challenging due to hallucination risks. Our pilot study (Section~\ref{sec:dsl}) confirms that LLMs struggle to directly generate complex DSL representations or concrete test scripts effectively.

To address these limitations, we propose {\tool}, an end-to-end \underline{T}r\underline{A}ffic \underline{R}ule-based test \underline{GE}nera\underline{T}ion framework. {\tool} leverages LLMs with in-context few-shot learning~\cite{liu2023visual} to extract information but avoids direct script generation by introducing a simple and expressive DSL. This DSL supports modular, compositional representations with syntactic and semantic validation to mitigate hallucination. Using validated DSL representations, {\tool} generates concrete test scenarios through templates and hierarchical map-based and dictionary-based route searches to populate parameters. The resulting scenarios enable ADS testing, with detailed recordings and reports identifying violations and issues.

We evaluated {\tool} on 284 scenarios derived from 54 traffic rules across seven ADSs: LAV~\cite{chen2022lav}, MMFN~\cite{zhang2022mmfn}, AUTO~\cite{carla_auto}, Autoware~\cite{kato2018autoware}, Apollo 7.0~\cite{Apollo}, Inverse Dynamics Model (IDM)~\cite{christiano2016transfer}, and Proximal Policy Optimization (PPO)~\cite{schulman2017proximal}. {\tool} uncovered 610 erroneous behaviors, including traffic rule violations and collisions. Common issues included failure to stop at stop signs or maintain safe distances. Autoware and Apollo had perception errors, MMFN faced localization inaccuracies, and LAV's cautious driving led to stalling. Two issues were confirmed by developers, and another matched known bug reports.

In summary, this work makes three key contributions:

\begin{itemize}
    \item We propose an end-to-end framework for automated test generation based on traffic rules, enabling violation detection and fault analysis for ADS testing. The code is open-sourced at \url{https://zenodo.org/records/14346539}.
    
    \item We are the first to use an LLM to extract knowledge from natural language traffic rules. To address LLM limitations in ADS test generation, we introduce a tailored DSL and a traffic rule parsing pipeline combining modular, in-context few-shot learning and knowledge validation for accurate scenario representation.

    \item We evaluate {\tool} on CARLA ,LGSVL and MetaDrive simulation platforms, generating 284 scenarios from 54 traffic rules to test seven ADSs. These tests uncovered 610 errors in five ADSs, with detailed logs facilitating root cause analysis.
\end{itemize}

The rest of the paper is organized as follows: Section~\ref{sec:background} introduces the motivation of this work. Section~\ref{sec:method} describes the proposed framework. Section~\ref{sec:experiment} outlines research questions and experiment settings. Section~\ref{sec:result} presents the results. Section~\ref{sec:related_work} compares related works. Section~\ref{sec:threat} discusses threats to validity. Section~\ref{sec:conclusion} summarizes the findings and suggests future research directions.

\section{Motivation}
\label{sec:background}
\begin{figure}
    \centering
    \includegraphics[width=1\linewidth]{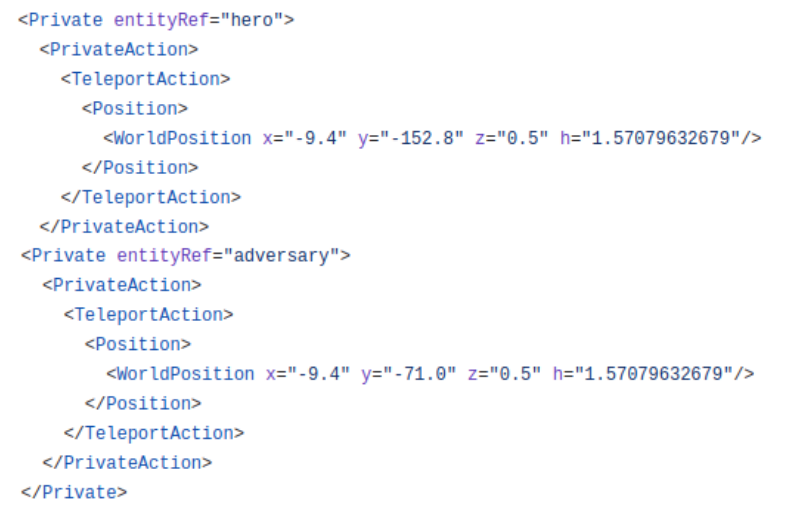}
    \caption{An example of parameter setup in OpenSCENARIO}
    \label{fig:openscenario_param}
\end{figure}

Menzel et al.~\cite{menzel2018scenarios} categorized ADS test scenarios into functional (abstract natural language descriptions), logical, and concrete scenarios (detailed with specific parameters). Similarly, Nalic et al.~\cite{nalic2020scenario} outlined knowledge-driven and data-driven testing approaches that progress from functional to logical and concrete scenarios. The NHTSA report~\cite{national2018functional} follows this methodology, starting with functional scenarios in natural language and refining them into logical and concrete simulations. Simulation testing often utilizes DSLs like OpenSCENARIO~\cite{openScenario} and Scenic~\cite{Scenic}, which are syntactically complex, verbose, and map-dependent. For example, a lane-changing scenario requires 75 lines in Scenic~\footnote{\href{https://github.com/BerkeleyLearnVerify/Scenic/blob/main/examples/carla/NHTSA_Scenarios/bypassing/bypassing_02.scenic}{Lane-changing scenario in Scenic syntax}} and 260 lines in OpenSCENARIO (Footnote~\ref{fn_dsl}). Moreover, these DSLs primarily describe concrete scenarios, requiring manual specification of critical parameters, such as initial vehicle positions, through iterative simulation experiments. Figure~\ref{fig:openscenario_param} illustrates parameter setup in OpenSCENARIO, highlighting the challenges of setting parameters for concrete scenarios in the test map.

LLMs show promise but struggle with low-granularity scenario generation due to the need for detailed map-based parameter searches, for which LLMs are not trained. 
Our pilot experiments (Section~\ref{sec:dsl}) revealed that LLM-generated Scenic and OpenSCENARIO scripts often contained syntax and parameter errors. These issues arise from the verbosity and complexity of existing DSLs, which increase the input context for the LLM, thereby exacerbating hallucination risks~\cite{liu2024lost}, as well as from the limitations of LLMs in handling map-based data.
To address these issues, we propose a modular, semantically rich, yet syntactically simple DSL. Each module generates concise prompts to reduce the length of input context~\cite{liu2024lost, glm2024chatglm} and mitigate hallucination risks~\cite{xu2024hallucination,chiang2022overcoming,huang2023survey}. Outputs undergo syntactic and semantic validation to ensure consistency. For low-granularity parameters, we use a program synthesis approach with pre-prepared templates for each simulator, leaving key parameters empty. These parameters are populated using dictionary-based and hierarchical map-based route searches, which our experiments show to be highly effective in generating accurate and consistent scenarios. This approach bridges the gap between traffic rules and simulator-based scenario generation, enabling robust and efficient ADS testing while addressing the limitations of existing DSLs and LLM-based solutions.

\section{Methodology}
\label{sec:method}

\subsection{Overview}
\begin{figure*}[h]
\centering
\includegraphics[width=1\textwidth]{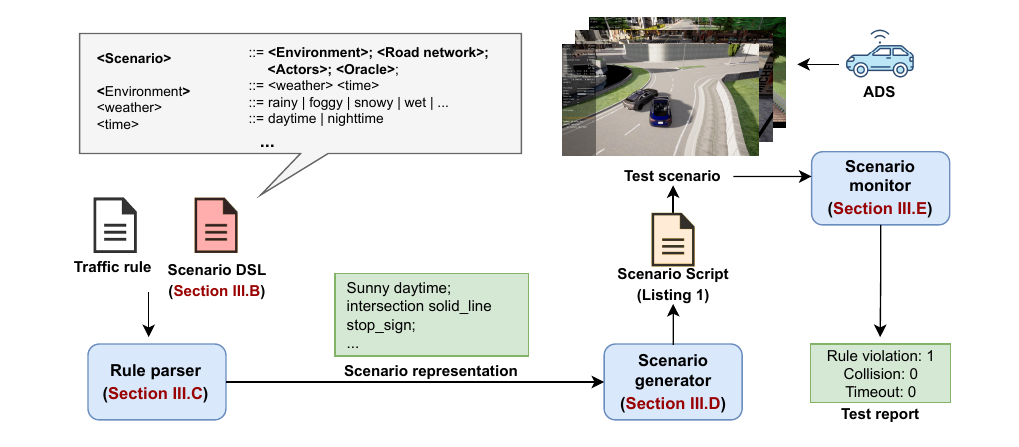}
\caption{An overview of the proposed test generation method}
\label{fig:overview}
\end{figure*}

Figure~\ref{fig:overview} illustrates the workflow of the proposed framework {\tool}. {\tool} takes three phases to parse a traffic rule description to an executable driving scenario in a simulator. The first phase is to parse the input traffic rule to a machine-readable test scenario representation, which follows the syntax of a proposed functional scenario-level DSL (Section~\ref{sec:dsl}). The rule parsing process is implemented in an LLM-powered \textit{rule parser}~(Section~\ref{sec:rule_parser}). In the second phase, the \textit{scenario generator} first creates a template scenario script. It then utilizes dictionary search and hierarchical map-based route search to identify precise parameters for functions in the template scenario script, forming the concrete test scenario script~(Section \ref{sec:generator}).
Lastly, the simulator executes the generated scenario script, rendering the corresponding test scenario with the ADS under test controlling the ego vehicle. Simultaneously, the \textit{Scenario monitor} captures the scenario execution process as a test scenario recording (video) and logs the simulation data, such as the positions and speeds of the ego vehicle and NPC actors. This logged data is used to identify the erroneous behaviors of the ADS under test and form a relevant test report, while the test scenario recording visually helps verify the ADS's erroneous behaviors and pinpoint their causes~(Section \ref{sec:test_monitor}).



\subsection{Domain Specific Language}
\label{sec:dsl}

Several DSLs, such as OpenSCENARIO, GeoScenario, and Scenic~\cite{openScenario, geoscenario, Scenic}, describe logical or concrete driving scenarios, but are syntactically complex and require manual parameter specification, limiting LLMs' ability to generate scenarios directly from traffic rules (Section~\ref{sec:background}). To address this, we propose a novel DSL for functional scenarios with simplified syntax, eliminating the need for concrete parameter values. This design enables modular learning and scenario generation for LLMs, reducing hallucination risks by avoiding lengthy prompts ~\cite{xu2024hallucination, chiang2022overcoming} and precise numerical parameter generation. Our pilot study results in Table~\ref{tab:pilot_study} demonstrate that existing DSLs, such as Scenic and OpenSCENARIO, lead to frequent syntax and parameter errors when used directly with LLMs. In contrast, the proposed DSL mitigates these issues, facilitating accurate and consistent scenario generation without requiring low-level parameter specifications.

\begin{table}[]
\scalebox{0.8}{
\begin{tabular}{l|c|c|l}
\hline
                                                                                    & \textbf{\begin{tabular}[c]{@{}c@{}}Number of parsed \\ traffic rules\end{tabular}} & \textbf{\begin{tabular}[c]{@{}c@{}}Number of \\ Executable  scenarios\end{tabular}} & \multicolumn{1}{c}{\textbf{Errors}}                                  \\ \hline
\textbf{\begin{tabular}[c]{@{}l@{}}Scenic scripts\end{tabular}} & 13                                                                           & 0                                                                            & \begin{tabular}[c]{@{}l@{}}Syntax Error\\ Attribute Error\end{tabular} \\ \hline
\textbf{\begin{tabular}[c]{@{}l@{}}CARLA Scenario \\ Runner scripts\end{tabular}}         & 8                                                                            & 0                                                                            & \begin{tabular}[c]{@{}l@{}}Function Name Error\\ Syntax Error\end{tabular}      \\ \hline
\end{tabular}}

\caption{Pilot study results of applying GPT-4 to directly generate Scenic and CARLA Scenario Runner scripts}
\label{tab:pilot_study}
\end{table}




\begin{figure*}[tb!]
\centering
{
$\def\arraystretch{1.1}
\setlength{\arraycolsep}{1pt}
\begin{array}{lll}
\textbf{\em \textless{}Scenario\textgreater{}} &::= \text{\em \textless{}Environment\textgreater{}}; \ \text{\em \textless{}Road network\textgreater{}}; \\ \ & \ \ \ \ \ \ \text{\em \textless{}Actors\textgreater{}}; \ \text{\em \textless{}Oracle\textgreater{}}; \\
\\
\textbf{\em \textless{}Environment\textgreater{}} &::= \text{\em \textless{}weather\textgreater{}} \ \text{\em \textless{}time\textgreater{}} \\
\text{\em \textless{}weather\textgreater{}} &::= \texttt{rainy} \ | \ \texttt{foggy} \ | \ \texttt{snowy} \ | \ \texttt{wet} \ | \ \texttt{...} \\
\text{\em \textless{}time\textgreater{}} &::= \texttt{daytime} \ | \ \texttt{nighttime} \\
\\
\textbf{\em \textless{}Road network\textgreater{}} &::= \text{\em \textless{}road type\textgreater{}} \ \text{\em \textless{}road marker\textgreater{}} \ \text{\em \textless{}traffic signs\textgreater{}} \  \\
\text{\em \textless{}road type\textgreater{}} &::= \ \texttt{intersection} \ | \ \texttt{roundabout} \ | \ \texttt{...} \\
\text{\em \textless{}road marker\textgreater{}} &::= \texttt{solid line} \ | \ \texttt{broken line} \ | \ \texttt{...} \\
\text{\em \textless{}traffic signs\textgreater{}} &::= \epsilon \ | \ \text{\em \textless{}traffic sign\textgreater{}}, \ \text{\em \textless{}traffic signs\textgreater{}} \\
\text{\em \textless{}traffic sign\textgreater{}} &::=  \texttt{stop sign} \ | \ \texttt{speed limit sign} \ | \ \texttt{...} \\
\\
\textbf{\em \textless{}Actors\textgreater{}} &::= \text{\em \textless{}ego vehicle\textgreater{}}, \  \text{\em \textless{}npc actors\textgreater{}} \\
\text{\em \textless{}ego vehicle\textgreater{}} &::= \text{\em \textless{}type\textgreater{}} \ \text{\em \textless{}behavior\textgreater{}} \ \text{\em \textless{}position\textgreater{}} \\
\text{\em \textless{}npc actors\textgreater{}} &::= \epsilon \ | \ \text{\em \textless{}npc actor\textgreater{}}, \ \text{\em \textless{}npc actors\textgreater{}} \\
\text{\em \textless{}npc actor\textgreater{}} &::= \text{\em \textless{}type\textgreater{}} \ \text{\em \textless{}behavior\textgreater{}} \ \text{\em \textless{}position\textgreater{}} \\
\text{\em \textless{}type\textgreater{}} &::= \texttt{car} \ | \ \texttt{truck} \ | \ \texttt{train} \ | \ \texttt{...} \\
\text{\em \textless{}behavior\textgreater{}} &::= \texttt{go forward} \ | \ \texttt{turn left} \ | \ \texttt{static} \ | \ \texttt{...} \\
\text{\em \textless{}position\textgreater{}} &::= \text{\em \textless{}position reference\textgreater{}} \ \text{\em \textless{}position relation\textgreater{}} \\
\text{\em \textless{}position reference\textgreater{}} &::= \texttt{ego vehicle} \ | \ \text{\em \textless{}road type\textgreater{}} \ | \ \text{\em \textless{}road marker\textgreater{}} \ | \ \text{\em \textless{}traffic sign\textgreater{}} \\
\text{\em \textless{}position relation\textgreater{}} &::= \texttt{front} \ | \ \texttt{behind} \ | \ \texttt{left} \ | \ \texttt{on} \ | \ \texttt{...} \\
\\
\textbf{\em \textless{}Oracle\textgreater{}} &::= \text{\em \textless{}longitudinal oracles\textgreater{}} \ \text{\em \textless{}lateral oracles\textgreater{}} \\
\text{\em \textless{}longitudinal oracles\textgreater{}} &::= \epsilon \ | \ \text{\em \textless{}longitudinal oracle\textgreater{}} \ \text{\em \textless{}longitudinal oracles\textgreater{}} \\
\text{\em \textless{}longitudinal oracle\textgreater{}} &::= \texttt{yield} \ | \ \texttt{decelerate} \ | \ ... \\
\text{\em \textless{}lateral oracles\textgreater{}} &::= \epsilon \ | \ \text{\em \textless{}lateral oracle\textgreater{}} \ \text{\em \textless{}lateral oracles\textgreater{}} \\
\text{\em \textless{}lateral oracle\textgreater{}} &::= \texttt{keep lane} \ | \ \texttt{change lane to left} \ | \ \texttt{...}
\end{array}$
}
\caption{The structure of Scenario DSL and a subset of elements in all components} 
\label{fig:schema}
\end{figure*}

Figure~\ref{fig:schema} shows the proposed DSL in the syntax of context-free grammar. A functional test scenario is composed of four primary components--- \textbf{\textit{Environment}},  \textbf{\textit{Road network}}, \textbf{\textit{Actors}}, and \textbf{\textit{Oracle}}. Each component comprises distinct subcomponents designed to delineate specific semantics.

\begin{itemize}
\item \textbf{\textit{Environment}} represents when the test scenario occurs, what the current \textit{weather} and \textit{time} are.

\item \textbf{\textit{Road network}} characterizes the geographical context of the test scenario, defining the \textit{road type}, and the presence of the \textit{road marker} and \textit{traffic signs}.

\item \textbf{\textit{Actor}} describes dynamic objects involved in the test scenario including the \textit{ego vehicle} and \textit{NPC actors}. For each actor, its \textit{type}, \textit{position}, and current \textit{behavior} when the scenario occurs are defined. Specifically, \textit{position} is composed of two subcomponents \textit{position reference} and \textit{position relation} to describe the relative position of the actor (the ego vehicle or the NPC actor) regarding the road network subcomponent elements or the ego vehicle when the actor is an NPC actor.

\item \textbf{\textit{Oracle}} regulates the appropriate behaviors of the ego vehicle in the \textit{longitudinal} direction (speed-related behaviors) and the \textit{lateral} direction (steering angle-related behaviors) during testing.
\end{itemize}


For each \textit{subcomponent}, its concrete value is called~\texttt{element} in this work. We define a list of optional elements for each subcomponent, derived from the traffic rule handbook~\cite{texasDriverHandbook} and OpenXOntology~\cite{openxontology}. OpenXOntology is designed by the Association for Standardization of Automation and Measuring Systems (ASAM)~\cite{asam}, which is an organization for developing standards for autonomous driving simulation and testing. OpenXOntology is the foundation ontology knowledge base that covers concepts and properties of elements that commonly occur in driving scenarios. Therefore, we chose to refer to it and the traffic rule handbook to define elements in the proposed DSL.


Figure 4 shows an example of a lane-changing scenario written in the proposed DSL, which is associated with the traffic rule ``You should keep a safe distance between your car and the one in front of you'', with the variation that the other vehicle is changing lanes to position itself in front of the ego vehicle. Compared with the same \href{https://github.com/carla-simulator/scenari_runner/blob/master/srunner/examples/LaneChangeSimple.xosc}{scenario} written in OpenSCENARIO as introduced in Section ~\ref{sec:background} and Scenic (Footnote~\ref{fn_dsl}), our proposed DSL describes the functional scenario at a higher level of abstraction without requiring precise parameter values. While OpenSCENARIO demands explicit definition of specific numeric parameters (such as exact vehicle coordinates shown in Figure~\ref{fig:openscenario_param}) and Scenic requires the implementation of programming language-like functions with verbose parameters, our DSL allows scenarios to be specified using abstract concepts that can be directly derived from traffic rules. This abstraction enables automatic scenario generation from rule descriptions and provides greater flexibility for instantiating scenarios across different simulation environments.


\begin{figure}
    \centering
    \begin{lstlisting}[language=Python, basicstyle=\ttfamily\small, style=customkeywords]
    Environment:
    Road_network: two-lane road
    Actors:
        @ego_vehicle@:
            @behavior@: go_forward
            @position_reference@: two-lane road
            @position_relation@: right
        @other_actors@:
            @actor_type@: car
            @behavior@: change_lane_to_right
            @position_reference@: ego_vehicle
            @position_relation@: left
    Oracle:
        @longitudinal_oracles@: decelerate
\end{lstlisting}
    \caption{An example of a lane-changing scenario described in the proposed DSL}
    \label{fig:example_scenario_dsl}
\end{figure}

\subsection{LLM-based Traffic Rule Parsing}
\label{sec:rule_parser}
In {\tool}, we propose using an LLM as a rule parser to convert traffic rules into scenario representations following the proposed DSL syntax. This approach leverages the recent advancements in LLMs, which have demonstrated human-level or superior performance in various NLP tasks~\cite{laskar2023systematic}. As the example shown in Table~\ref{tab:prompt_example}, the LLM, GPT-4~\cite{ChatGPT}, clearly explains the meaning of the given traffic rule in the prompt. It even infers the item \textit{``T'' intersection} that is not explicitly mentioned in the traffic rule and shows some keywords in bold. Inspired by this, we design a systematic pipeline for traffic rule parsing using an LLM.

However, the utilization of LLMs presents several challenges:
\begin{itemize}
    \item Prompt Engineering: The primary method of engaging LLMs is through prompts, which are simple cues or instructions guiding the model's output. Designing effective prompts is crucial for obtaining accurate and relevant responses.
    \item Hallucination: A significant challenge in using LLMs is the hallucination problem~\cite{zhang2023siren, ye2023cognitive}, where models may generate incorrect or fabricated information not present in the input data. This is particularly critical when dealing with traffic rules, where accuracy is paramount.
\end{itemize}

\begin{figure*}
    \centering
    \includegraphics[width = .95\textwidth]
    {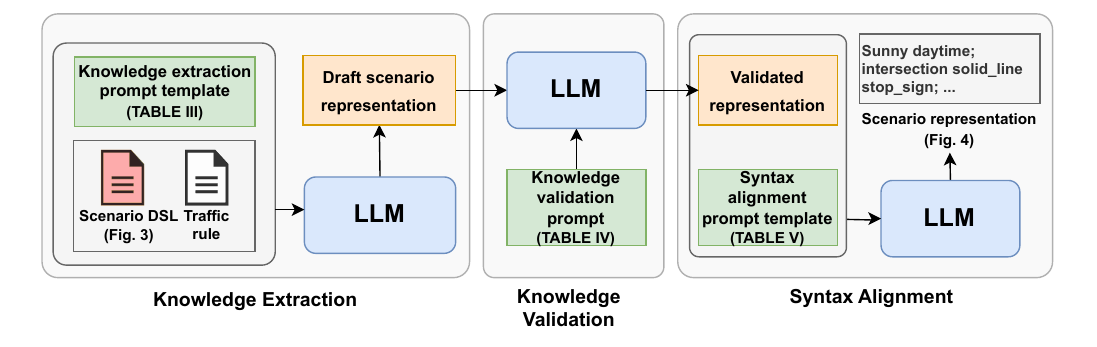}
    \caption{The workflow of traffic rule parsing}
    \label{fig:enter-label}
\end{figure*}

To address these challenges, we propose a novel pipeline (illustrated in Figure~\ref{fig:enter-label}) that overcomes the limitations of standard LLM usage. Our approach consists of three key components:
\begin{itemize}
    \item Few-shot In-context Learning-based Knowledge Extraction: This method improves the LLM's ability to extract relevant information from traffic rules by providing it with examples of correct extractions.
    \item Knowledge Validation: To mitigate the hallucination problem, we implement a validation step to verify the extracted information against the original traffic rule.
    \item Syntax Alignment: This step ensures that the extracted scenario representations are consistently formatted and aligned with the input traffic rules.
\end{itemize}

Each of these components is guided by carefully designed prompts, addressing the challenge of effective prompt engineering. This comprehensive approach ensures that the LLM extracts accurate scenario representations that are consistent with input traffic rules, overcoming the primary challenges of LLM utilization in this context.

\begin{table}[h!]

\begin{tabular}{p{0.95\columnwidth}}
\hline
\rowcolor[HTML]{C0C0C0} 
\textbf{Prompt}                                                                                                                                                                                                                                                                                                                                                                                                                                            \\
\begin{tabular}[c]{@{}l@{}}Briefly explain the traffic rule: When approaching an intersection of a \\ through street traveling from a street that ends at the intersection, you must \\ stop and yield the right-of-way to vehicles on the through street.\end{tabular}                                                                                                                                                                      \\
\rowcolor[HTML]{C0C0C0} 
\textbf{GPT-4 Response}                                                                                                                                                                                                                                                                                                                                                                                                                                    \\
\begin{tabular}[c]{@{}l@{}}When you're driving on a street that comes to an end at a through street \\(often called a ``T'' intersection), you need to:\\ 1. \textbf{Stop} before reaching the intersection.\\ 2. \textbf{Yield the right-of-way}: This means you must wait and let vehicles on \\ the through street go first. They have priority.\\ 3. \textbf{Proceed safely}: Once it's safe to do so, you can continue your journey.\end{tabular} \\ \hline
\end{tabular}
\caption{An example of interacting with the LLM GPT-4 using the prompt}
\label{tab:prompt_example}
\end{table}





\subsubsection{Knowledge Extraction}
In the first step, we design the prompts following the in-context few-shot learning paradigm~\cite{liu2023visual}. As illustrated in Table~\ref{tab:prompt_knowledge_extraction}, initially, we establish the role of the LLM as a domain expert in autonomous driving, followed by a brief introduction of the task. This role-setting enables the LLM to better identify the related knowledge it needs to produce high-quality replies. In the primary prompt, we first present the specifics of the proposed DSL's definition and the list of optional elements within the DSL to guide the LLM in comprehending the task's context. Then, we provide a traffic rule-scenario representation pair as a learning example. The example enables the LLM to learn the specific structures of the task's input and output. Finally, we provide the input traffic rule to the LLM for knowledge extraction and generating a draft scenario representation. The detailed content of the prompt can be found in the supplementary material~\footnote{Supplementary material: \url{https://shorturl.at/gFPX3}}. The few-shot in-context learning lets the LLM learn the syntax of our proposed DSL and provides examples for it to learn the correct output format of the scenario representation.    

\begin{table}[h!]
\begin{tabular}{p{0.95\columnwidth}}
\hline
\rowcolor[HTML]{C0C0C0} 
\textbf{Role Setting}                                                                                                                                                                                                                                                                                                                                                                                                                                                                                                                                                                                                                                                                                                                                                                                                                             \\
\begin{tabular}[c]{@{}l@{}}You are a test expert for autonomous driving systems. Your task is to \\ generate specific test scenario representations from given traffic rules.\end{tabular}                                                                                                                                                                                                                                                                                                                                                                                                                                                                                                                                                                                                                                              \\
\rowcolor[HTML]{C0C0C0} 
\textbf{Prompt}                                                                                                                                                                                                                                                                                                                                                                                                                                                                                                                                                                                                                                                                                                                                                                                                                           \\
\begin{tabular}[c]{@{}l@{}}Below is the definition of a domain-specific language to represent test \\ scenarios for autonomous driving systems:\\ \{\textit{Details of DSL}\}\\ \\ Below are the lists of commonly used elements for each subcomponent. \\ When creating the scenario representation, consider the following \\ elements first for each subcomponent. If no element can describe the \\ close meaning, create a new element by yourself.\\ \{\textit{Details of element lists}\}\\ \\ Below is an example of an input traffic rule and the corresponding \\ scenario representation: Traffic rule: \{\textit{content of the traffic rule}\}\\ Scenario representation: \{\textit{content of the scenario representation}\} \\ \\ Based on the above descriptions and examples, convert the following \\ traffic rule to corresponding scenario representation: \\ \{\textit{input traffic rule}\} \end{tabular} \\ \hline
\end{tabular}
\caption{The prompt template for knowledge extraction}
\label{tab:prompt_knowledge_extraction}
\end{table}

\subsubsection{Knowledge Validation}
With the generated scenario representation in the knowledge extraction, we then validate whether the extracted knowledge and generated scenario representation are correct. Recent works~\cite{miao2023selfcheck, helbling2023llm, manakul2023selfcheckgpt} found that LLM has the ability to self-check whether its output is correct and consistent with the given inputs. Therefore, we construct the prompt as displayed in Table~\ref{tab:prompt_knowledge_revision} to guide the LLM to verify whether the extracted component elements in the draft scenario representation align with the input traffic rule. In cases of inconsistency, the LLM is directed to make revisions and generate a validated scenario representation.


\begin{table}[h]
\begin{tabular}{p{0.95\columnwidth}}
\hline
\rowcolor[HTML]{C0C0C0} 
\textbf{Prompt}              
\\
\begin{tabular}[c]{@{}l@{}}Are the elements in the generated scenario description consistent with the \\ input traffic rule? If not, correct the inconsistencies and output the revised \\ scenario representation.\end{tabular} \\                                                                                                                                                                           
\rowcolor[HTML]{C0C0C0} 
\textbf{LLM Response}                                                                                                                                                                                                                   \\
\{\textit{Content of the LLM response}\}                                                                   \\              
\hline
\end{tabular}
\caption{The prompt for knowledge validation}
\label{tab:prompt_knowledge_revision}
\end{table}

\subsubsection{Syntax Alignment}

When the validated scenario representation is obtained, the final step is to align it with the syntax of the proposed DSL. This is done to ensure that each subcomponent's element in the scenario representation corresponds with the element in the predefined list but not other words with similar semantic meanings.  For instance, in the case of actor behavior, the LLM's output might be ``go straight" instead of the "go forward" defined in the element list. To solve the problem, as demonstrated in Table~\ref{tab:prompt_alignment}, the LLM is guided to select the element in the element list defined in the DSL that has the closest meaning to update the LLM's output element in the scenario representation. Furthermore, to maintain the rule parser's scalability for undefined elements, such as newly introduced traffic signs in traffic rules, the LLM is also directed to retain its original response if it fails to identify a similar element in the element list.

\begin{table}[]

\begin{tabular}{p{0.95\columnwidth}}
\hline
\rowcolor[HTML]{C0C0C0} 
\textbf{Prompt}                                                                                                                                                                                                                                                                                                                                                            \\
\begin{tabular}[c]{@{}l@{}}For each element \{\textit{content of the LLM's output element}\} of subcomponent \\\{\textit{subcomponent name}\}, find out an element with the closest meaning to \\your output element from the following element list. If all elements in the \\element list do not express a similar meaning as your output element, keep\\ your output element as the answer.\\ \{\textit{Content of the element list in the DSL}\}\end{tabular} \\
\rowcolor[HTML]{C0C0C0} 
\textbf{LLM Response}                                                                                                                                                                                                                                                                                                                                                      \\
Content of the LLM response                                                                                                                                                                                                                                                                                                                                                \\ \hline
\end{tabular}
\caption{The prompt template for syntax alignment}
\label{tab:prompt_alignment}
\end{table}

\subsection{Template-based Test Script Generation}
\label{sec:generator}

After the functional scenario description is generated by the rule parser, it is then used to construct a concrete scenario that can run in simulators. Inspired by templated-based code generation~\cite{jorges2013state}, we propose a template-based test script generation method to do this. We first create a \textit{template script} that outlines the core functions required to execute a test scenario using the simulator's APIs, as shown in Listing~\ref{lst:template_script}. The key challenge is determining the concrete values for parameters such as \textit{weather parameter}, \textit{time parameter}, \textit{ego\_initial\_coordiate}, \textit{npc\_initial\_coordinate}, \textit{ego\_destination\_coordinate}, \textit{npc\_destination\_coordinate}, \textit{longitudinal oracle}, and \textit{lateral oracle}. These parameters must be derived from the abstract scenario representation and simulator-specific information to create an executable test script.

To illustrate our approach, consider a traffic rule: ``When approaching intersections not controlled by signs or signals, yield the right-of-way to any vehicle that has entered or is approaching the intersection on your right.'' The rule parser would generate a scenario representation containing elements like 
    \begin{verbatim}
    Environment:
        Weather: clear 
        Time: day 
    Road_network:
        road_type: intersection 
        traffic_signs: none 
    Actors:
        ego_vehicle:
            behavior: go_forward 
        other_actors: 
            actor_type: vehicle
            behavior: go_forward
            position_reference: intersection
            position_relation: right
    Oracle: 
        longitudinal_oracle: yield
    \end{verbatim}
    
Our generator must transform this abstract representation into concrete parameter values for the template script.

We propose two complementary approaches to identify these parameter values: \textit{dictionary search} and \textit{Hierarchical Map-based Route Search}. The overall process is shown in Algorithm~\ref{alg:param}.

\begin{lstlisting}[language=Python, caption=Simplified template scenario script,label={lst:template_script},captionpos=b]
def run_scenario(*args):
    set_environment(weather_params, time_params)
    set_actors(ego_initial_coordinate, 
               ego_destination_coordinate,
               npc_initial_coordinate, 
               npc_destination_coordinate)
    set_test_oracle(longitudinal_oracle, 
                    lateral_oracle)

def set_environment(weather_params, time_params):
    simulator.set_weather(weather_params)
    simulator.set_time(time_params)

def set_actors(ego_initial_coordinate, 
               ego_destination_coordinate,
               npc_initial_coordinate, 
               npc_destination_coordinate):
    simulator.set_actor_position(ego_vehicle, 
                      ego_initial_coordinate)
    ads.move_to(ego_destination_coordinate)
    
    simulator.set_actor_position(npc_actor, 
                    npc_initial_coordinate)
    simulator.move_npc_to(npc_actor,
                npc_destination_coordinate)

def set_test_oracle(longitudinal_oracle,
                            lateral_oracle):
    monitor.set_oracles(longitudinal_oracle,
                             lateral_oracle)

 \end{lstlisting}

\begin{algorithm*}
\SetAlgoLined
\SetKwInput{Input}{Input~}
\SetKwInput{Output}{Output~}
    \SetKwData{Scenario}{SP}
    \SetKwData{ParamDict}{PD}
    \SetKwData{Params}{PV} 
    \SetKwData{MapInfo}{MapInfo} 
    \SetKwData{Simulator}{Simulator} 
    \SetKwData{Route}{Route}
    \SetKwData{Routes}{Routes}
    \SetKwData{QR}{QR}
    \SetKwData{NearbyRoutes}{NearbyRoutes}
    \SetKwData{EgoRoutes}{EgoRoutes}
    \SetKwData{EgoRoute}{EgoRoute}
    \SetKwData{NPCRoute}{NPCRoute}
    \SetKwData{Flag}{Flag}
    
    \SetKwFunction{GetMapInfo}{GetMapInfo}
    \SetKwFunction{GetRoutes}{GetRoutes}
    \SetKwFunction{Waypoints}{Waypoints}
    \SetKwFunction{GetPrevRoute}{GetPrevRoute}
    \SetKwFunction{FilterRouteByRoadNetwork}
    {FilterRouteByRoadNetwork}
    \SetKwFunction{Append}{append}
    \SetKwFunction{FilterRouteByActor}{FilterRouteByActor}
    \SetKwFunction{SampleWaypoints}{SampleWaypoints}
    \SetKwFunction{CheckEgoRoute}{CheckEgoRoute}
    \SetKwFunction{CheckNPCRoute}{CheckNPCRoute}
    \SetKwFunction{GetNearbyRoutes}{GetNearbyRoutes}

    \Input{\Scenario: test scenario representation in dictionary format; \ParamDict: Lookup table of parameter values for subcomponents Environment and Oracle}
    \Output{\Params: dictionary of parameter values}
    \ccomment{ \tcc{Get parameter values for parameters weather\_params, time\_params, longitudinal\_oracle, and lateral\_oracle by dictionary search}}
    \Params[``weather''] $\leftarrow$ \{\Scenario[``Environment''][``Weather'']\; \ParamDict[\Scenario[``Environment''][``Weather'']]\}\; 
    \Params[``time''] $\leftarrow$ \ParamDict[\Scenario[``Environment''][``Time'']]\;
    \Params[``longitudinal\_oracle''] $\leftarrow$ \ParamDict[\Scenario[``Oracle''][``longitudinal\_oracle'']]\;
    \Params[``lateral\_oracle''] $\leftarrow$ \ParamDict[\Scenario[``Oracle''][``lateral\_oracle'']]\;        
    
    \ccomment{\tcc{Initialize \Routes and \QR to save all routes and filtered routes that match \Scenario[``Road\_network'']}}
    \Routes $\leftarrow$ []\;
    \QR $\leftarrow$ []\ ;
    \MapInfo $\leftarrow$ \GetMapInfo(\Simulator)\;
    \Waypoints $\leftarrow$ \SampleWaypoints(\Simulator)\;
    \Routes $\leftarrow$ \GetRoutes(\Waypoints, \MapInfo)\;
    
    \ForEach {\Route $\in$ \Routes}{%
        \If {\FilterRouteByRoadNetwork(\Route, \Scenario[``Road\_network''])}{
            \QR.\Append{\Route}\;
        }
    }

    \ForEach {\Route $\in$ \QR}{%
        \If {\FilterRouteByActor(\Route, \Scenario[``Actors''][``ego\_vehicle''])}{
            \EgoRoutes.\Append{\Route}\;
        }
    }
    
    \ForEach {\Route $\in$ \EgoRoutes}{%
        \NearbyRoutes $\leftarrow$ \GetNearbyRoutes(\Route)\;
        \ForEach {\NPCRoute $\in$ \NearbyRoutes}{
            \If {\FilterRouteByActor(\NPCRoute, \Scenario[``Actors''][``other\_actors''])}{
                \Flag $\leftarrow$ \textsf{true}\;
                \textsf{break}\;
            }            
        }
        \If {\Flag}{
            \textsf{break}\;
        }
    }
    
    \ccomment{ \tcc{Get parameter values for parameters ego\_initial\_coordinate, ego\_destination\_coordinate, npc\_initial\_coordinate, npc\_destination\_coordinate}}
    \If {\Flag}{
        \Params[``ego\_initial\_coordinate''] $\leftarrow$ \GetPrevRoute(\EgoRoute).start\_point\;
        \Params[``ego\_destination\_coordinate''] $\leftarrow$ \EgoRoute.end\_point\;
        \Params[``npc\_initial\_coordinate''] $\leftarrow$ \NPCRoute.start\_point\;
        \Params[``npc\_destination\_coordinate''] $\leftarrow$ \NPCRoute.end\_point\;
        
    }
    
\caption{Scenario Parameter Identification}
\label{alg:param}
\end{algorithm*}

\begin{figure}[tb!]
\centering
\includegraphics[width = .5\textwidth]{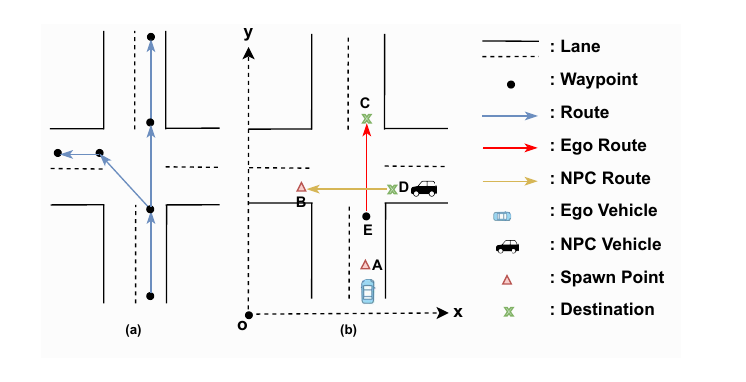}
\caption{The demonstration of waypoints, routes, and a driving scenario}
\label{fig:route_scenario}
\end{figure}

\subsubsection{Dictionary Search for Environment and Oracle Parameters} 
Lines 2-5 in Algorithm~\ref{alg:param} show how to obtain parameter values for weather\_params, time\_params, longitudinal\_oracle, and lateral\_oracle through a dictionary search. We create a predefined lookup dictionary \texttt{PD} that maps abstract concepts from the scenario representation to concrete simulator-specific parameter values.

For example, in our implementation for the Carla simulator, the abstract weather condition ``clear'' in the scenario representation is mapped to concrete simulator parameters like \texttt{\{fog\_density: 0, rain\_density: 0\}} in the parameter dictionary \texttt{PD}. Similarly, the time parameter ``day'' is mapped to \texttt{\{hour: 12, minute: 0\}} in Carla. For test oracles, an abstract oracle like ``yield'' is mapped to specific monitoring functions (such as \texttt{GiveWayTest} in Carla) that check if the ego vehicle reduces speed appropriately when approaching an intersection.

This dictionary-based approach allows us to maintain a clean separation between the abstract scenario representation and the simulator-specific implementation details. The dictionaries are created once for each supported simulator and can be extended to support additional environmental conditions or oracle types.

\subsubsection{Hierarchical Map-based Route Search for Actor Coordinates} 
\label{concrete_function_Search} 
The most challenging parameters to determine are the precise coordinates for the ego vehicle and NPC actors. Unlike environmental parameters that can be directly mapped, these coordinates must be identified based on the road network structure of the specific simulation map and the behavioral constraints specified in the scenario representation.

We propose a hierarchical map-based route-search algorithm that follows four key steps: \textit{Routes Generation}, \textit{Route Filtering}, \textit{Ego Route Identification}, and \textit{Actors Coordinate Identification}, as shown in Lines 8 to 40 in Algorithm~\ref{alg:param}.

\textbf{Routes Generation}: First, our algorithm obtains comprehensive map information from the simulator, including the waypoints (coordinates) of all lanes and traffic signs. We implement a uniform sampling algorithm to extract waypoints along all lanes within the map. \textit{Routes} are then constructed by establishing connections between adjacent sampled waypoints along each lane, as illustrated in Figure~\ref{fig:route_scenario} (a). For each generated \textit{route}, we compute a set of attributes based on the map information, including route length, direction, connecting routes, presence of traffic signs, and whether the route traverses an intersection or other road types.

For example, in Figure~\ref{fig:route_scenario} (b), a coordinate system is established, which defines the origin point at the bottom left corner. The route from E to C contains information that the start waypoint is at coordinates (5, 5), the end waypoint is at (5, 10), and the route direction is straight.

\textbf{Route Filtering}: With all routes identified, we filter them based on the road network constraints specified in the scenario representation. For our intersection example, we filter for routes that pass through intersections without traffic signs or signals, as specified in the scenario representation. The filtered routes are stored in the list \texttt{QR}, as shown in Lines 13 to 15 in Algorithm~\ref{alg:param} 

\textbf{Ego Route Identification}: Next, we analyze the filtered routes to identify suitable candidates for the ego vehicle based on the behavioral constraints specified in the scenario representation. For our example, we look for routes where the ego vehicle can ``go forward'' through an intersection, such as the route from E to C in Figure~\ref{fig:route_scenario} (b). This process yields a set of candidate ego routes stored in \texttt{EgoRoutes}.

\textbf{NPC Route Identification}: For each candidate ego route, we examine nearby routes to find suitable paths for NPC actors that satisfy their behavioral constraints. In our intersection example, we need to find a route for an NPC vehicle approaching from the right of the ego vehicle, such as the route from B to D in Figure~\ref{fig:route_scenario} (b). If we find a valid combination of ego and NPC routes, we designate them as the final routes. If no valid combination can be found after examining all candidates, we declare the scenario infeasible for the current simulation map.

\textbf{Actors Coordinate Identification}: Once the ego and NPC routes are identified, we determine the concrete coordinate parameters. The ego vehicle's initial position (\texttt{ego\_initial\_coordinate}) is set at the start waypoint of the route preceding the ego route (e.g., at point A in Figure~\ref{fig:route_scenario} (b) with coordinates (5, 2)), allowing the ADS to accelerate to a steady speed before reaching the intersection. The destination (\texttt{ego\_destination\_coordinate}) is set to the end waypoint of the ego route (point C with coordinates (5, 10)). For the NPC vehicle, the initial position (\texttt{npc\_initial\_coordinate}) is set to the start of its route (point D with coordinates (6, 6)), and the destination (\texttt{npc\_destination\_coordinate}) is set to the end of its route (point B with coordinates (2, 6)).

\textbf{Executable Script Formation}: Once all parameter values are automatically identified, they are filled into the template scenario script to create an executable test script that can be directly loaded by simulators like Carla. Then the ADS under test such as Autoware can be integrated into the test scenario to control the ego vehicle driving in the scenario. Unlike traditional DSLs such as OpenSCENARIO or Scenic, where testers must manually pre-define precise actor coordinates and trajectories, our approach derives these values automatically from abstract scenario representations. This automation significantly reduces the effort required to create test scenarios and eliminates the tight coupling between scenario descriptions and specific simulation maps. The same abstract scenario can be instantiated across different maps without manual adjustments, enabling more efficient and flexible testing of ADS.

\subsection{Scenario Monitor}\label{sec:test_monitor}


Once the test scenario script is executed by the simulator, the scenario monitor initiates the recording of the test scenario, counts the length of running time for the test scenario, and logs essential scenario data. Once the ego vehicle controlled by the ADS reaches the destination, the scenario monitor saves the test scenario recording, scenario running time, and scenario data. This data includes various parameters such as actor positions, speeds, and steer angles, recorded for each frame. If the ADS becomes stuck and cannot reach the destination, the scenario monitor also handles scenario termination when the running time reaches a predetermined timeout limit (60 seconds). This limit is calculated based on route lengths and actor speeds, ensuring sufficient time for a properly functioning ADS to reach the destination.

The test scenario recording is used to check the behavior of the ADSs by our predefined test oracle functions and help analyze the underlying problems of the ADS. The logged data is used to generate the test report regarding the following three aspects: (1) \texttt{rule violation}, whether the ADS violates the test oracles; (2) \texttt{collision}, whether the ADS collides with NPC actors or roadside objects; (3) \texttt{timeout}, whether the ADS reaches the destination within the given time limit.

For identifying \texttt{rule violation}, in~{\tool}, we have implemented test oracle functions for all possible longitudinal oracle elements (e.g., \texttt{yield}) and lateral oracle elements (e.g., \texttt{keep lane}) defined in the DSL. These test oracle functions take the logged scenario data as input and respond to whether the ADS under test violates the corresponding test oracle. For example, the test oracle function for the lateral oracle \textit{keep lane} will check whether the ADS does not change to other lanes based on position data. \texttt{Collision} and \texttt{Timeout} are checked by the position data of all actors and the duration of the saved test scenario recording respectively.

\section{Experiment}
\label{sec:experiment}


\subsection{Research Questions}
We proposed three research questions (RQs) and conducted experiments to evaluate the effectiveness of {\tool}. The first RQ investigates the ability of the rule parser. The second and the 
last 
RQs evaluate the performance of {\tool} for generating test scenarios and detecting erroneous behaviors of ADSs.


\begin{itemize}

\item RQ1: How effectively can LLM-powered \textit{Rule Parser} generate test scenario representations from traffic rules?

\item RQ2: Can {\tool} produce executable test scenarios that comply with traffic rules?


\item RQ3: Can {\tool} accurately identify incorrect ADS behavior?

\end{itemize}

\subsection{Benchmark}
\subsubsection{Traffic Rule Benchmark}
\label{sec:benchmark}
To evaluate these RQs, we first created a traffic rule benchmark based on Texas Driver Handbook~\cite{texasDriverHandbook}. The driver handbook has 14 chapters in total, in which Chapters 4-9 introduce traffic rules for the safe driving of vehicles. To create the benchmark, we selected sentences from Chapters 4-9 based on the following criteria:

Initially, we excluded traffic rules with accompanying figures as current general-purpose language models can't process multimodal data. Next, we removed rules on non-driving scenarios, like parking, due to limited ADS support. Finally, rules regarding non-vehicular behaviors were omitted, as our focus is on ADSs. Following the criteria, we filtered 18 rules illustrated with figures, 16 rules about non-driving scenarios, and 8 rules related to non-vehicular behaviors. After filtering, we retained 98 traffic rules detailed in the supplementary material~\footnote{Supplementary material: \url{https://shorturl.at/gFPX3}}. For each traffic rule, we used the DSL to create a ground truth scenario representation. All authors made these independently, reviewed each other's work, and reached a consensus to ensure fidelity to the rule's content.

\subsubsection{Test Scenario Benchmark}
We tried to generate test scenarios for each traffic rule in the simulators Carla~\cite{dosovitskiy2017carla} and LGSVL~\cite{rong2020lgsvl} because they are widely used in ADS testing. Specifically, following the settings in ADS testing researches~\cite{deng2022scenario, tian2022mosat}, we used eight maps (Town 1-7, and Town 10) in Carla and four maps in LGSVL (San Francisco, Shalun, Gomentum, and Cubetown) to generate test scenarios. As described in Table~\ref{tab:maps}, these maps are representative to cover driving environments in various rural, urban, high way areas, and some of them are digital twins reconstructed from real-world testing environments for ADS.
These maps contain most of typical road networks, such as single and multi-lane roads, intersections, and roundabouts, for generating corresponding testing scenarios.
However, the original version of Carla contains limited traffic signs in the maps.
In order to expand Carla's applicability to a broader range of scenarios, we incorporated additional assets, such as traffic signs and railway tracks, into the simulator maps. Consequently, our customized version of the Carla simulator successfully supports scenario generation for $54$ out of the total $98$ traffic rules, supporting more than twice of traffic rules as the original version of Carla. On the other hand, as LGSVL has ceased official development and there are limited instructions, we did not add new assets and used original maps and assets in LGSVL. Consequently, $40$ rules are supported to generate test scenarios in LGSVL.

\begin{table}[]
\scalebox{0.9}{
\begin{tabular}{ll}
\hline
\rowcolor[rgb]{0.804,0.804,0.804} \textbf{Map}  & \textbf{Description}                                                                                                                    \\ 
Town 01       & A small, simple town with a river and several bridges.                                                                                  \\
\rowcolor[rgb]{0.898,0.898,0.898} Tonw 02       & \begin{tabular}[c]{@{}l@{}}A small simple town with a mixture of residential and\\ commercial buildings.\end{tabular}                   \\
Town 03       & A larger, urban map with a roundabout and large junctions.                                                                              \\
\rowcolor[rgb]{0.898,0.898,0.898} Town 04       & \begin{tabular}[c]{@{}l@{}}A small town embedded in the mountains with a special \\ "figure of 8" infinite highway.\end{tabular}        \\
Town 05       & \begin{tabular}[c]{@{}l@{}}Squared-grid town with cross junctions and a bridge. It \\ has multiple lanes per direction.\end{tabular}    \\
\rowcolor[rgb]{0.898,0.898,0.898} Town 06       & \begin{tabular}[c]{@{}l@{}}Long many lane highways with many highway entrances\\ and exits.\end{tabular}                                \\
Town 07       & \begin{tabular}[c]{@{}l@{}}A rural environment with narrow roads, corn, barns and \\ hardly any traffic lights.\end{tabular}            \\
\rowcolor[rgb]{0.898,0.898,0.898} Town 10       & \begin{tabular}[c]{@{}l@{}}A downtown urban environment with skyscrapers, residential\\  buildings and an ocean promenade.\end{tabular} \\
San Francisco & A digital twin of urban environment from San Francisco, CA.                                                                             \\
\rowcolor[rgb]{0.898,0.898,0.898} Shalun        & A digital twin of Taiwan Car Lab testing environment.                                                                                   \\
Gomentum      & A digital twin of Gomentum ADS testing environment.                                                                                     \\
\rowcolor[rgb]{0.898,0.898,0.898} Cubetown      & \begin{tabular}[c]{@{}l@{}}A urban environment containing intersections, traffic lights, \\ and traffic signs.\end{tabular}             \\ \hline
\end{tabular}}
\caption{Description of maps used in experiments}
\label{tab:maps}
\end{table}

In experiments, we used all rules to evaluate the performance of \textit{Rule Parser} in RQ1 and used the simulator-compatible rules in the simulators for RQ2 and RQ3. For each traffic rule within the supported rules, {\tool} attempted to generate driving scenarios on eight maps (Town 1-7, and Town 10) in Carla and four maps in LGSVL (San Francisco, Shalun, Gomentum, and Cubetown). As described in Table~\ref{tab:maps}, these maps present driving environments in various rural and urban areas, containing most of typical road networks, traffic signs and road markers.

As different maps encompass different traffic signs and road markers, a particular traffic rule may not necessarily generate a test scenario across all maps. Nevertheless, for each rule, we sought to generate corresponding test scenarios across all compatible maps. In total, $217$ and $124$ test scenarios were generated in Carla and LGSVL respectively.

To further verify the generalizability of {\tool}, we integrated {\tool} with another simulator called Metadrive~\cite{li2022metadrive}, which mainly focuses on supporting the development and test of reinforcement learning (RL)-based autonomous driving agents. By implementing the template scenario scripts in Metadrive, the test generation pipeline can seamlessly integrate with the simulator. Specifically, we generated $27$ test scenarios in Metadrive for supported traffic rules.

\subsubsection{ADSs under Test}
In RQ2, we tested four ADSs including Auto~\cite{carla_auto}, MMFN~\cite{zhang2022mmfn}, LAV~\cite{chen2022lav}, and Autoware~\cite{kato2018autoware} on all generated driving scenarios in Carla and one ADS Apollo 7.0~\cite{Apollo} in LGSVL. Auto~\cite{carla_auto} is an ADS modified based on the NPC agent in Carla. MMFN and LAV are two ADSs that achieved good performances on the Carla leaderboard challenge~\cite{carla_challenge}. These three ADSs were obtained from the Github repositories\footnote{MMFN and Auto: \url{https://github.com/Kin-Zhang/carla-expert}}~\footnote{LAV: \url{https://github.com/dotchen/LAV}}. We used their pre-trained models with default settings. Autoware~\cite{kato2018autoware} is a ROS-based industrial-level ADS and is extensively used for testing ADS~\cite{feng2023dense}~\cite{zhang2022finding}. We tested Baidu Apollo 7.0~\cite{Apollo} in LGSVL because it is an industrial-level ADS and widely used in prior works. In Metadrive, we tested two built-in RL-based autonomous driving policies Inverse Dynamics Model(IDM) policy~\cite{christiano2016transfer} and Proximal policy optimization(PPO) policy~\cite{schulman2017proximal}.

\subsubsection{LLM Used in Rule Parser}
In this work, we adopted GPT-4~\cite{ChatGPT} as the knowledge extraction model in \textit{Rule Parser}. We chose to use GPT-4 because it is the current state-of-the-art LLM that is widely known and easy to access.  We used the official API provided by OpenAI to invoke GPT-4 following the proposed rule parsing pipeline. To evaluate the effectiveness of the proposed rule parsing pipeline (Knowledge extraction -\textgreater{} Knowledge Validation -\textgreater{} Syntax Alignment) in Section~\ref{sec:rule_parser}, we created four versions of the rule parser using GPT-4, namely \textit{GPT-4-KE}, \textit{GPT-4-KE(-FL)} \textit{GPT-4-KE-KV}, and \textit{GPT-4-KE-KV-SA} for ablation study,  which correspond to only applying knowledge extraction, applying knowledge extraction without few-shot in-context learning, applying knowledge extraction and validation, applying knowledge extraction, knowledge validation, and syntax alignment respectively.
In addition, we also evaluated and compared the performance of the rule parser powered by different LLMs, including GPT-3, GPT-3.5, and two versions of a state-of-the-art open-source LLM Llama3.1 8B~\cite{llama} and Llama3.1 70B. We directly run the llama models via an open source software Ollama~\cite{ollama} without fine-tuning on a machine with an RTX4090 GPU. The detailed experiment setting is introduced in Section~\ref{sec:exp_rq1}.

\subsection{Experiment Settings}

\subsubsection{RQ1}
\label{sec:exp_rq1}
For each traffic rule in the benchmark, we applied the rule parser powered by different LLMs to generate the corresponding scenario representation. We then compared the generated scenario representation with the ground truth to evaluate the effectiveness of the rule parsers. 

We compared our work with LawBreaker~\cite{sun2022lawbreaker}, which uses traffic laws to guide the fuzzing of test scenarios. As their work manually parses traffic rules for fuzzing pre-defined test scenarios, while our goal is to generate test scenarios accurately by automatically parsing traffic rules, it is difficult to conduct a fair, end-to-end comparison with LawBreaker. Therefore, we only evaluated the rule parsing ability of {\tool} to check whether {\tool} can automatically parse traffic rules used in LawBreaker and then generate the corresponding test scenarios.





\subsubsection{RQ2}
Given the traffic rules that can be rendered to test scenarios in the simulator Carla, we run {\tool} to generate test scenarios. During the process, we saved the running outputs of {\tool}, including \textit{compile errors} (errors that are thrown by the program compiler such as the Python compiler to compile the test scenario script to the executable test scenario), \textit{runtime errors} (errors thrown by the simulator to terminate the running test scenario), and \textit{test scenario recordings} if no error happened during the execution of the test scenario. The log of compile errors and runtime errors was used to evaluate whether {\tool} can generate executable test scenarios based on traffic rules. The recorded test scenarios were evaluated to determine their accuracy in adhering to the described traffic rules.

To evaluate the accuracy of the generated test scenarios, 
we conducted a human study, which has been validated that human evaluation results can align with safety metrics in a recent study~\cite{birchler2024does}. Informed consent was obtained from all participants involved in the study. Given our large set of 54 traffic rules and their implementation across multiple simulation maps, manually reviewing all generated scenarios would be prohibitively time-consuming and costly. Therefore, we randomly sampled 18 out of 54 rules, resulting in 75 scenarios in total for human evaluation, to ensure a manageable yet representative assessment. The selected rules are listed in the supplementary material~\footnote{Supplementary material: \url{https://shorturl.at/gFPX3}}.

We conducted two human evaluation studies. First, we divided the $75$ test scenarios into five online surveys, randomly given to software engineering students across five practical testing classes at a public university. Participant numbers were 36, 15, 27, 16, and 10 across the five classes for each survey due to class sizes and willingness to participate. Participants in the survey only needed to check the consistency between traffic rules and scenario recording, such as whether a car stops at a stop sign in a scenario recording as described in the traffic rule. 

In addition, considering that students may not have sufficient driving experience, we published the five surveys on a researcher-orientated crowdsource platform Prolific~\cite{prolific} to hire workers to evaluate generated test scenarios. We set the requirement that each participant in the study should have a driving license and driving experience. Each survey on Prolifc is finished by 20 participants.



The links to online surveys can be found in the supplementary material. We asked participants to evaluate the accuracy of generated scenarios considering the corresponding descriptions in the traffic rules, using a five-point scale that included \textit{Totally match}, \textit{Mostly match}, \textit{Partially match}, \textit{Mostly not match}, and \textit{Totally not match}. A detailed explanation of the above options is provided in the supplementary material. In the human study, each participant was shown the traffic rule along with the corresponding generated driving scenario in the Carla simulator. By viewing the driving scenario video, they were then required to evaluate the scenario's adherence to the traffic rule description.



\subsubsection{RQ3} For each ADS under test, we integrate it into all generated test scenarios to control the movement of the ego vehicle. During the execution of test scenarios, test reports and running logs are recorded for further analysis of erroneous behavior of the ADS.



\begin{table*}[]
\scalebox{0.95}{
\begin{tabular}{ccccccccc}
\hline
\rowcolor[rgb]{0.804,0.804,0.804} \textbf{Subcomponent} & \textbf{GPT-4-KE-SV-SA} & \textbf{GPT-4-KE-SV} & \textbf{GPT-4-KE} & \multicolumn{1}{l}{{ \textbf{GPT-4-KE(-FL)}}} & \textbf{GPT-3.5} & \textbf{GPT-3} & \multicolumn{1}{l}{{ \textbf{LLAMA3.1-8b}}} & \multicolumn{1}{l}{{ \textbf{LLAMA3.1-70b}}} \\
Weather               & \textbf{1.0}            & \textbf{1.0}         & \textbf{1.0}      & { \textbf{1.0}}                               & \textbf{1.0}     & \textbf{1.0}   & { \textbf{1.0}}                           & { \textbf{1.0}}                            \\
\rowcolor[rgb]{0.898,0.898,0.898} Time                  & \textbf{1.0}            & \textbf{1.0}         & \textbf{1.0}      & { \textbf{1.0}}                               & \textbf{1.0}     & \textbf{1.0}   & { \textbf{1.0}}                           & { \textbf{1.0}}                            \\
Road type             & \textbf{0.980}          & 0.969                & 0.959             & { 0.969}                                      & 0.908            & 0.949          & { 0.693}                                  & { 0.908}                                   \\
\rowcolor[rgb]{0.898,0.898,0.898} Road marker           & \textbf{1.0}            & \textbf{1.0}         & \textbf{1.0}      & { \textbf{1.0}}                               & \textbf{1.0}     & 0.959          & { 0.796}                                  & { \textbf{1.0}}                            \\
Traffic sign          & \textbf{1.0}            & \textbf{1.0}         & \textbf{1.0}      & { \textbf{1.0}}                               & \textbf{1.0}     & 0.990          & { 0.918}                                  & { 0.959}                                   \\
\rowcolor[rgb]{0.898,0.898,0.898} Type                  & \textbf{1.0}            & \textbf{1.0}         & \textbf{1.0}      & { \textbf{1.0}}                               & 0.944            & 0.964          & { 0.929}                                  & { 0.949}                                   \\
Behavior              & \textbf{0.980}          & 0.969                & 0.969             & { \textbf{0.969}}                             & 0.872            & 0.857          & { 0.755}                                  & { 0.827}                                   \\
\rowcolor[rgb]{0.898,0.898,0.898} Position reference    & \textbf{0.985}          & 0.934                & 0.918             & { 0.878}                                      & 0.745            & 0.827          & { 0.388}                                  & { 0.796}                                   \\
Position relation     & \textbf{0.985}          & 0.949                & 0.949             & { 0.918}                                      & 0.699            & 0.724          & { 0.316}                                  & { 0.684}                                   \\
\rowcolor[rgb]{0.898,0.898,0.898} Longitudinal oracle   & \textbf{1.0}            & \textbf{1.0}         & 0.990             & { \textbf{0.990}}                             & 0.857            & 0.878          & { 0.816}                                  & { 0.867}                                   \\
Lateral oracle        & \textbf{1.0}            & 0.980                & 0.980             & { \textbf{0.980}}                             & 0.980            & 0.959          & { 0.939}                                  & { 0.959}                                   \\ \hline
\end{tabular}}
\caption{Accuracies at the component level for different rule parsers}
\label{tab:parse_rule}
\end{table*}

\subsection{Evaluation Metrics}

\subsubsection{RQ1}

We used \textbf{Accuracy} to measure the performance of the rule parser. Specifically, we applied the metric at \textit{rule} level and \textit{component} level. Given a scenario representation of a traffic rule $R={R^1, R^2, ..., R^m}$ containing $M$ component elements, the \textbf{accuracy at the rule level} is calculated as Formula~\ref{eq:acc_rule}, where $GT$ is the ground truth scenario representation of the traffic rule generated as described in Section~\ref{sec:benchmark}. $ACC_{rule}=100\%$ means all component elements are parsed correctly for the traffic rule. For a rule parser, a higher number of traffic rules parsed with 100\% $ACC_{rule}$ indicates that it is more proficient in deriving information from traffic rules and creating structured scenario representations.
\begin{equation}
\label{eq:acc_rule}
    ACC_{rule} = \frac{\sum_{i=1}^{M}(R^i==GT^i)}{M}
\end{equation}

Given $N$ traffic rules ${T_1, T_2, ..., T_N}$ and the specific component $i$, the \textbf{accuracy at the component level} $ACC_{comp}(i)$ is calculated as Formula~\ref{eq:acc_comp}, where $R_j^i=$ is the parsed element for the $i$th component of the traffic rule $T_j$ and  $GT_j^i$ is the corresponding ground truth component element. The accuracy at the component level demonstrates the ability of the rule parser to process diverse information associated with different components.
\begin{equation}
\label{eq:acc_comp}
    ACC_{comp}(i) = \frac{\sum_{j=1}^{N}(R_j^i==GT_j^i)}{N}
\end{equation}




\subsubsection{RQ2}
We assessed {\tool}'s capability to generate executable test scenarios using three metrics: \textbf{Compile error rate}, \textbf{Runtime error rate}, and \textbf{Normal execution rate}.  These metrics respectively represent the frequencies of compile errors, runtime errors, and test scenario recordings collected from successful executions for all produced test scenario scripts.


Furthermore, to determine if the generated test scenarios align with traffic rules, we analyzed the \textbf{voting distribution} across five options for all test scenarios in the human study. To ensure the validity of the human study findings, we utilized weighted \textbf{Fleiss' Kappa}~\cite{fleiss1971measuring} to assess the voting agreements among participants. Specifically, we improved the original Fleiss' Kappa by incorporating linear weights into the agreement calculation process of Fleiss' Kappa following the instruction in~\cite{linearWeight}. Typically, the value of Fleiss' Kappa ranges between -1 and 1. A value less than 0 signifies poor agreement, while a value of 1 indicates perfect agreement~\cite{hallgren2012computing}.



\subsubsection{RQ3}
We summarized the performances of these ADSs based on testing reports generated by the scenario monitor in Section~\ref{sec:test_monitor}, which contains three metrics \textbf{rule violation}, \textbf{collision}, and \textbf{timeout}. The time limit of \texttt{timeout} was set as 60 seconds,  which is enough for the ADS under testing to reach the destination in generated test scenarios if the ADS drives normally.

\section{Results}
\label{sec:result}

\subsection{RQ1: Effectiveness of Rule Parser}

\begin{figure*}[tb!]
\centering
\includegraphics[width = .9\textwidth]{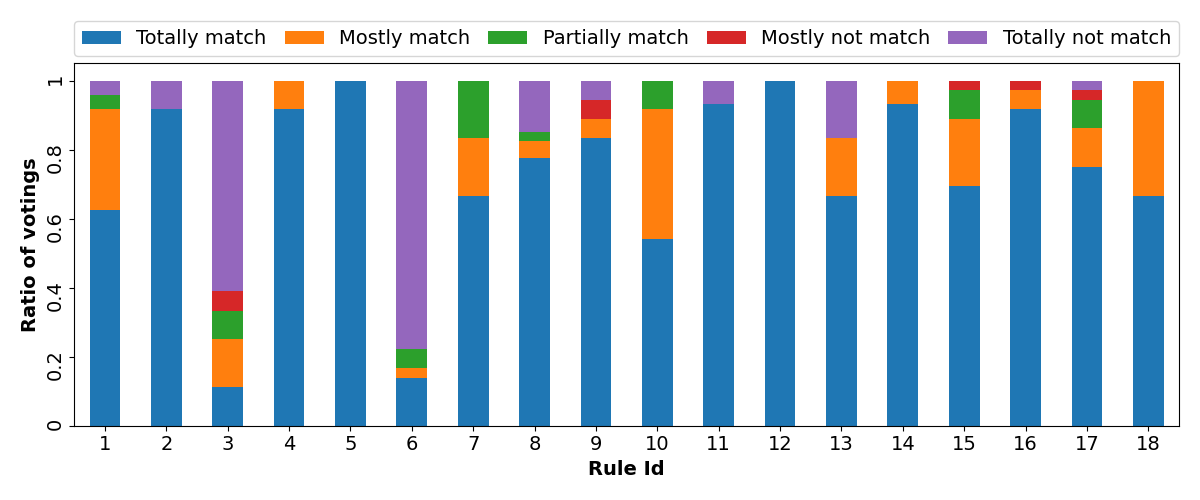}
\caption{Overall votings of scenarios on 18 rules by students}
\label{fig:pie_chart}
\vspace{-2mm}
\end{figure*}

\begin{figure*}[tb!]
\centering
\includegraphics[width = .9\textwidth]{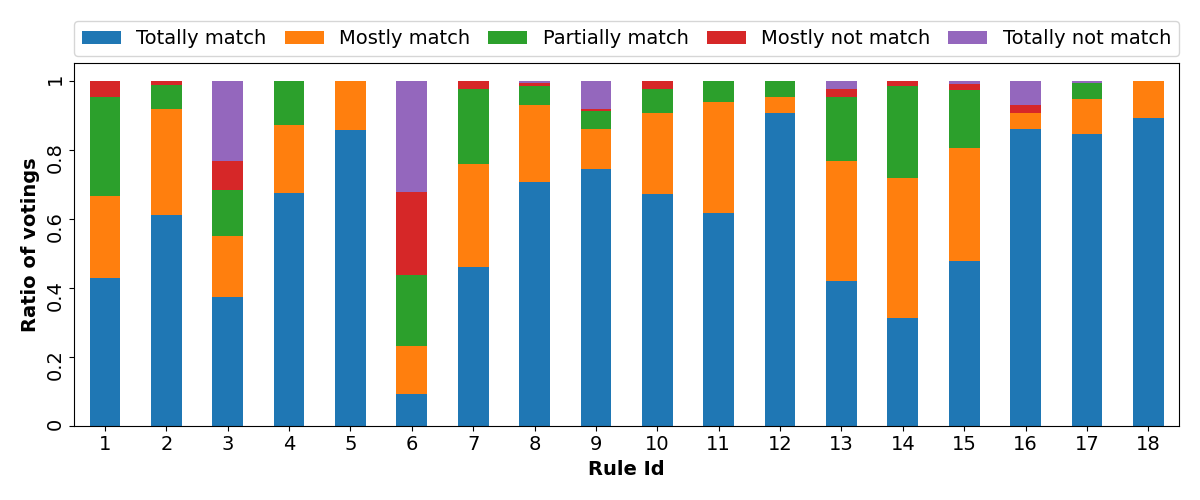}
\caption{Overall votings of scenarios on 18 rules by Prolific workers}
\label{fig:voting_2}
\vspace{-2mm}
\end{figure*}

We calculated rule-parsing accuracies on 98 rules in the benchmark for five rule parsers.
For GPT-4-KE, 71 rules can be parsed with 100\% accuracy. The number is further improved by GPT-4-KE-KV and GPT-4-KE-KV-SA, achieving 73 and 90 respectively. For GPT-4-KE-KV, only 2 rules have less than 90\% parsing accuracy. The results demonstrate that the proposed rule-parsing pipeline is effective in improving the rule-parsing accuracy. The performances of GPT-3.5 and GPT-3 are similar, with only about 30 rules parsed with 100\% accuracy. The result means that a more powerful LLM can significantly improve the performance of rule parsing.

Table~\ref{tab:parse_rule} shows the results of average component-parsing accuracy among all traffic rules in the benchmark for different rule parsers. From the table, all rule parsers can achieve 100\% accuracies on the subcomponents \textit{Weather} and \textit{Time}. The GPT-4-powered rule parsers can achieve more than 90\% accuracies on all subcomponents while GPT-3.5 and GPT-3 perform not well on the subcomponents \textit{Behavior}, \textit{Position reference}, \textit{position relation}, and \textit{Longitudinal oracle}. The result shows that rule parsers powered by GPT-4 have better logical inference abilities to infer actor behaviors and positions given traffic rules that do not contain accurate descriptions for these subcomponents. 

Compared with GPT-4-KE, GPT-4-KE-(-FL) performs similarly on most of subcomponents except \textit{Position reference} and \textit{Position relation}. The reason is that for the subcomponents like \textit{Weather} and \textit{Time}, most of the input traffic rules already have clear descriptions for them. Even without few-shot in-context learning, the LLM can still extract such information correctly. On the other hand, for subcomponents \textit{Position reference} and \textit{Position relation}, GPT-4-KE-(-FL) performs about 3\% worse than GPT-4-KE. This performance disparity underscores the significant potential of few-shot in-context learning to enhance the model's task comprehension and inference capabilities, particularly for more complex semantic components. However, it is crucial to note that for GPT-4-KE, the absence of syntax validation and alignment mechanisms persists as a critical limitation. Even with few-shot in-context learning, the model continues to experience hallucination challenges when attempting to extract precise position-related information.

LLAMA3-8b achieves the lowest accuracies on all components due to its limited parameter size, while LLAMA-70b achieves similar performance as GPT-3.5 on most of the components, which shows the potential to fine-tune open-source LLMs targeting ADS test generation tasks. Another future work direction is to explore multi-modality LLM to generate more accurate scenario representations based on the text and corresponding driving image.

We further investigated the wrong parsing results of GPT-4-KE-SV-SA. For \textit{Road type}, the rule parser failed to infer \texttt{multi-lane road} when the traffic rule only implicitly mentioned that the actors were changing lanes. For \textit{Behavior}, the rule parser is sometimes confused about the current driving behavior and the expected driving behaviors (i.e., \textit{longitudinal oracle} and \textit{lateral oracle}). For \texttt{Position reference} and \texttt{Position relation}, the rule parser occasionally chose the wrong reference frame for actors. For example, it may output \textit{Position reference} as \texttt{intersection} and \textit{Position relation} as \texttt{front} to demonstrate that the \texttt{intersection} is in the \texttt{front} of the ego vehicle, while in the proposed DSL the relation should be presented as the ego vehicle is \texttt{behind} the \texttt{intersection}. The possible reason is that the rule parser forgets a part of the information in the provided context about the proposed DSL. Overall, even though the rule parser powered by GPT-4 with the proposed rule-parsing pipeline performed well on most of the traffic rules in the benchmark, it still has some problems in inferring implicit information and forgetting context information. Future work can be conducted to explore how to solve the problems for better rule-parsing performance.

To compare with LawBreaker~\cite{sun2022lawbreaker}, we checked the traffic regulations used in their work. Specifically, 38 traffic rules in total are manually parsed in their work. We input these traffic rules into {\tool} to automatically generate the corresponding scenarios. 33 out of 38 traffic rules can be successfully parsed and generate the corresponding scenarios. The parsed scenario representations and generated scenarios can be found in the supplementary material~\footnote{Supplementary material: \url{https://shorturl.at/gFPX3}}. The reason for failing to parse the rest traffic rules is that these traffic rules describe multiple consecutive driving behaviors of vehicles, which are not supported in the current scenario representation DSL. We will leave this for future work to refine the proposed DSL.

 
\subsection{RQ2: Capability of Test Scenario Generation}



\subsubsection{Execution Statuses of Generated Test Scenarios}
For generated test scenarios, 90\% executed normally, while 7\% had compile errors, and 3\% had runtime errors. We found that most errors were caused by potential bugs in Carla. Specifically, incorrect waypoint coordinates from the CARLA API caused vehicles to initialize under the ground, leading to simulation termination. Compile errors arose when actor initial positions collided with existing map objects, resulting in ``None'' object initialization and subsequent API call issues. A similar bug has been reported by other contributors on Github~\footnote{Bug report about waypoint height: \url{https://github.com/carla-simulator/carla/issues/2323}}. In essence, {\tool} typically generates executable scenarios, but inherent CARLA bugs occasionally disrupt them.


\subsubsection{Accuracy of Generated Test Scenarios in Accordance with Traffic Rules}
For human studies finished by students and experienced workers, the voting results are generally similar. Within the set of 75 test scenarios voted for by students, 50 were perceived to be at least a partial match to the rules. None of these 50 scenarios received votes for the \textit{Most not match} or \textit{Totally not match} categories. For participants with driving experiences, 49 scenarios were voted without \textit{Most not match} or \textit{Totally not match} categories.

Figure~\ref{fig:pie_chart} reveals the voting result by students on all rules about whether the generated test scenarios match the descriptions of the traffic rules. Each bar represents the distribution of votes for the five categories, across all generated scenarios on different maps corresponding to a specific traffic rule. For the majority of test scenarios for the traffic rules, votes predominantly converged on the \textit{Totally match} or \textit{Mostly match} categories. This suggests that, on the whole, the generated scenarios are consistent with the articulated traffic rules.

\begin{figure}[htbp]
    \centering
    \begin{subfigure}[b]{0.45\textwidth}
        \centering
        \includegraphics[width=\textwidth]{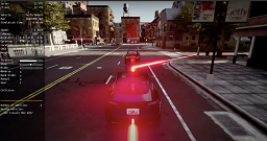}
        \caption{A generated scenario for unbroken line}
        \label{fig:broken}
    \end{subfigure}
    \hfill
    \begin{subfigure}[b]{0.45\textwidth}
        \centering
        \includegraphics[width=\textwidth]{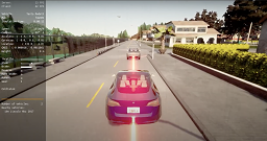}
        \caption{A generated scenario for wet weather}
        \label{fig:wet}
    \end{subfigure}
    \caption{Examples of scenarios that do not match with traffic rule descriptions}
    \label{fig:invalid_scenarios}
\end{figure}

Figure~\ref{fig:voting_2} reveals a similar voting trend among the rules for participants from the Prolific platform. Most \textit{Totally not match} and \textit{Mostly not match} votes appear in Rules 3 and 6, aligning with the student participants' results. Compared to student voting, the Prolific results show lower ratios of \textit{Totally match} and \textit{Totally not match} responses. This discrepancy may be attributed to two factors. First, Prolific participants might be more conservative in selecting extreme options, possibly due to concerns about reward eligibility if such responses are deemed invalid or filtered out by the survey publisher. Second, prolific workers may consider the entire driving scenario when voting. For instance, regarding Rule 6 (which states the ego vehicle should maintain a safe distance from the front car in wet weather), as shown in Figure~\ref{fig:wet}, students might vote \textit{Totally not match} if the wet weather is not properly rendered. In contrast, Prolific participants might choose \textit{Mostly not match} or \textit{Partially match} if the ego vehicle is indeed maintaining a safe distance from the front vehicle, despite the weather rendering issues.

To further analyze why participants thought some generated test scenarios do not match descriptions in the corresponding traffic rules, we manually checked test scenarios 
whose \textit{Totally match} ratio is under 60\%. Two factors were identified. First, the weather rendered by the simulator on some maps is low fidelity. For example, for the weather effect ``wet'', the rendered scenarios in some maps do not look wet, more like shadows, as shown in Figure~\ref{fig:wet}. Second, the road information about road marker obtained from Carla is not accurate. Most of the broken lines are annotated as solid lines, as an example shown in Figure~\ref{fig:broken} which causes the inconsistency of test scenarios that occur on the road with a solid line.


To evaluate the validity of the human study, we calculated Fleiss' Kappa~\cite{fleiss1971measuring} for five surveys separately and then obtained the mean Fleiss' Kappa. The Fleiss' Kappa values for five surveys finished by students are 0.59, 0.70, 0.67, 0.67, and 0.79 respectively. The values for surveys finished by Prolific workers are 0.58, 0.65, 0.68, 0.68, and 0.67 respectively. The mean Fleiss' Kappa value for students and workers are 0.68 and 0.67 respectively, which shows a substantial degree of consensus among the raters following the rating standard in~\cite{fleiss1971measuring}. In summary, {\tool} is believed to accurately generate test scenarios that match the corresponding traffic rules.

\subsection{RQ3: Rule Violation Detection on ADSs}

\begin{figure*}

\centering
\subfloat[Traffic rule violation (Auto model~\cite{carla_auto}): Not Give way to the other vehicle in roundabout]{\includegraphics[height=0.16\textwidth,width=.22\textwidth]{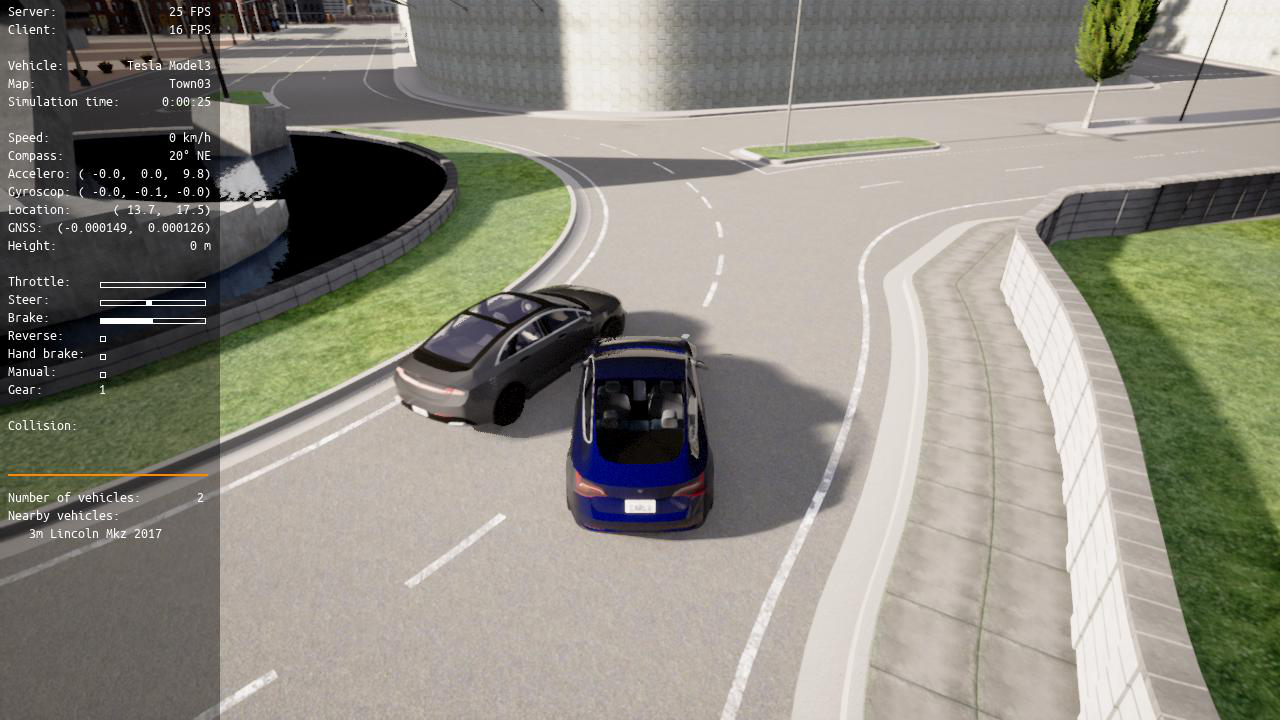}\label{fig:violation_roundabout}}\hfil
\subfloat[Traffic rule violation (Auto model~\cite{carla_auto}): Not give way to the other vehicle at an intersection]
{\includegraphics[height=0.16\textwidth,width=.22\textwidth]{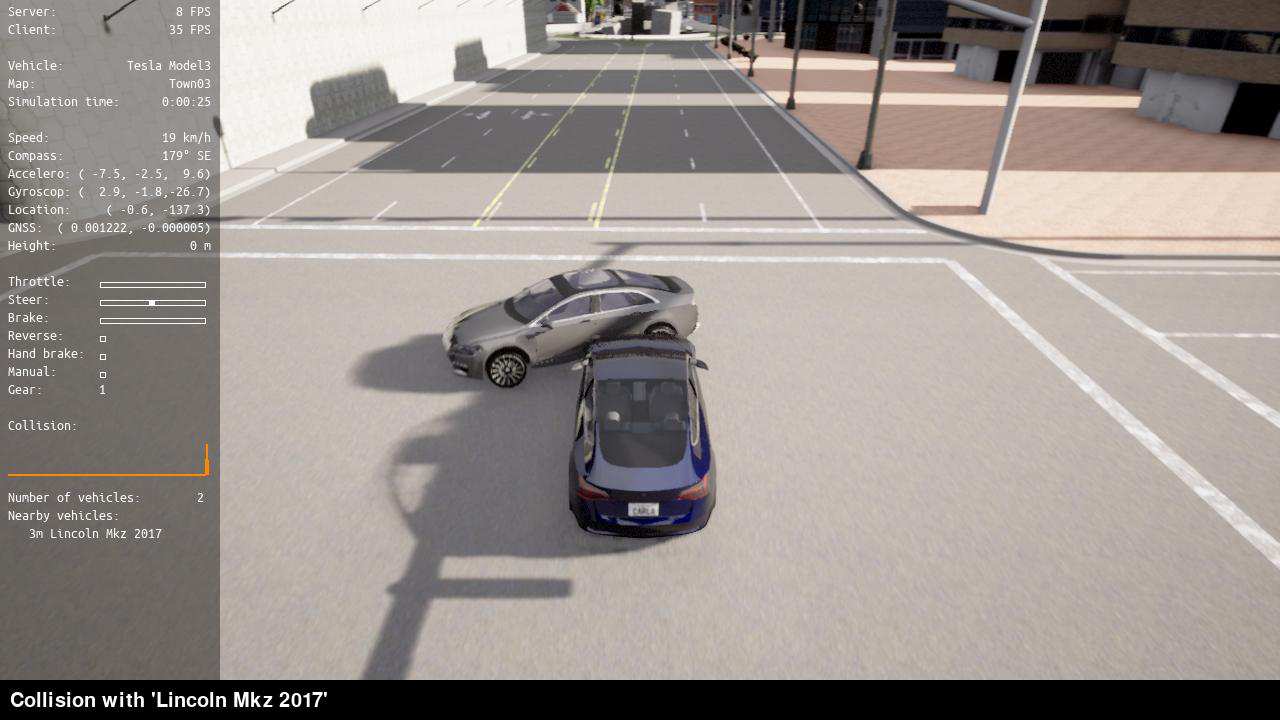}\label{fig:violation_intersection}}\hfil
\subfloat[Traffic rule violation (MMFN model~\cite{zhang2022mmfn}): Not stop before the stop sign]
{\includegraphics[height=0.16\textwidth,width=.22\textwidth]{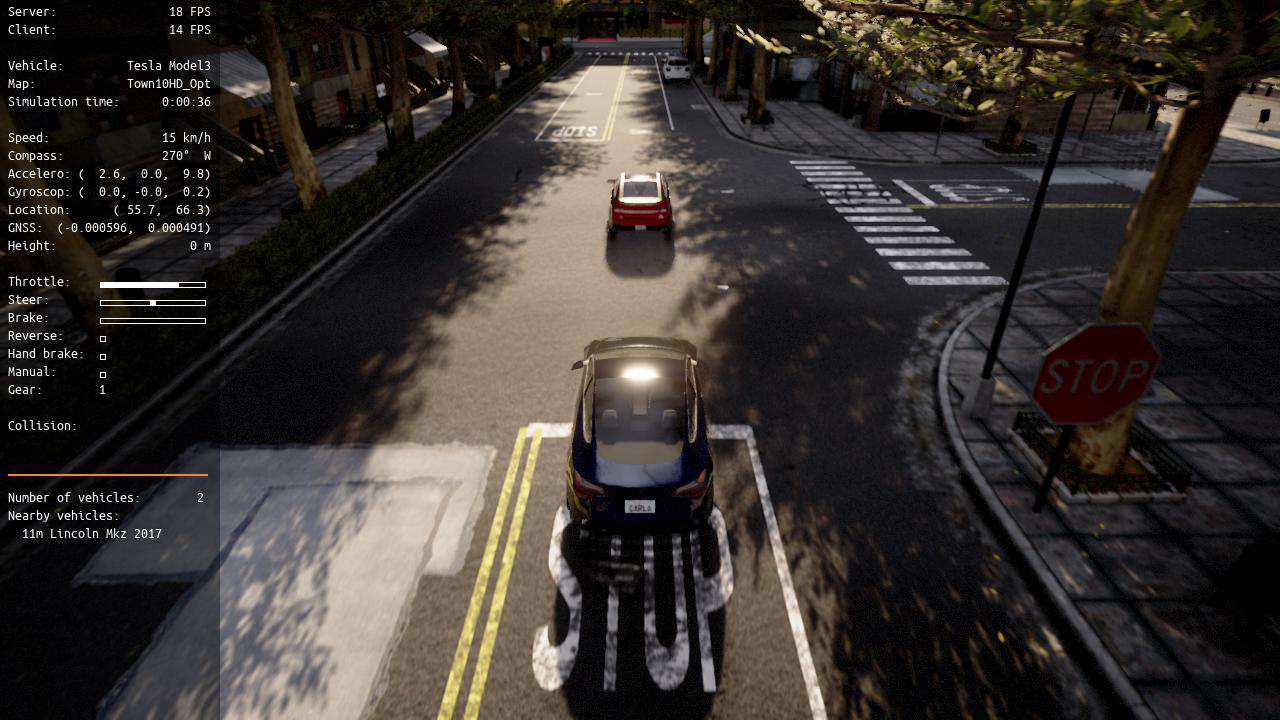}\label{fig:violation_stop}}\hfil
\subfloat[Traffic rule violation (Apollo~\cite{Apollo}): Collision with a pedestrian crossing the road]
{\includegraphics[height=0.16\textwidth,width=.22\textwidth]{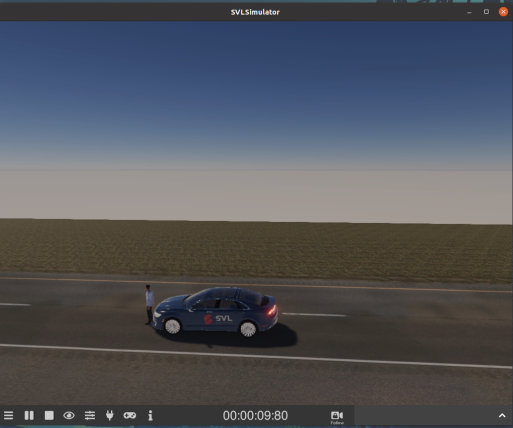}\label{fig:violation_collision_apollo}}\hfil

\subfloat[MMFN model~\cite{zhang2022mmfn} collides on the roadside]{\includegraphics[height=0.16\textwidth,width=.22\textwidth]{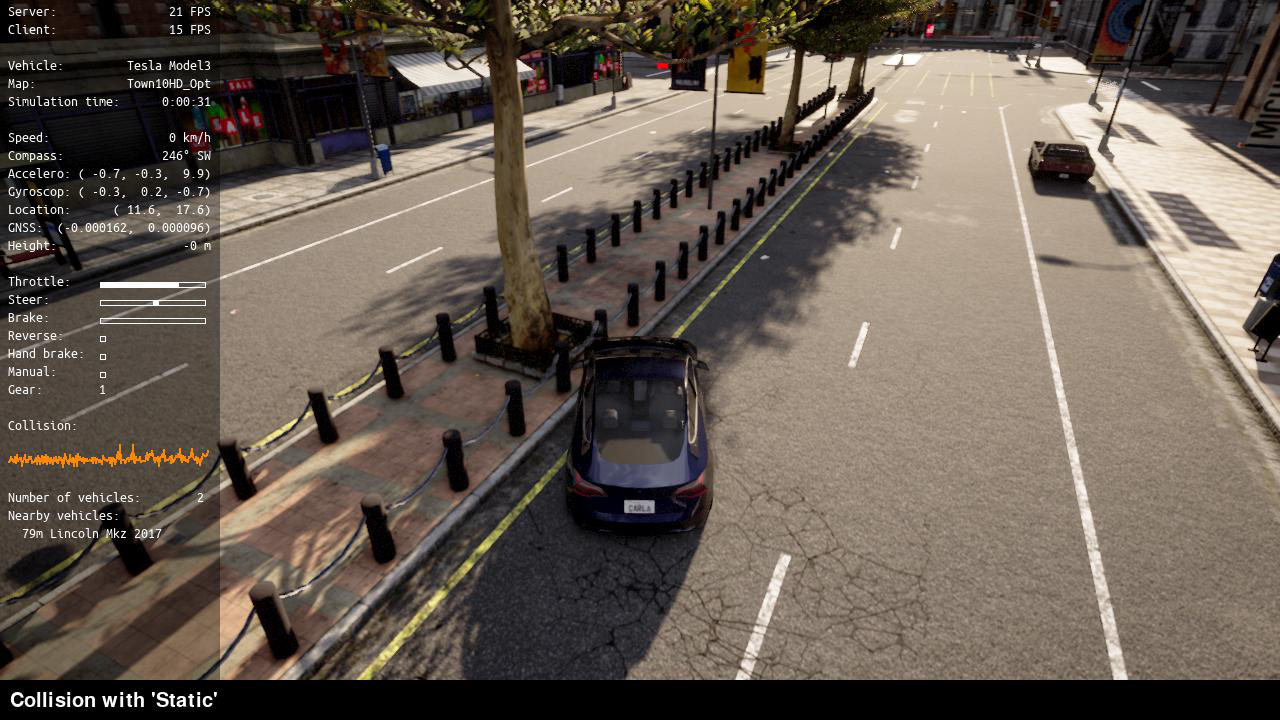}\label{fig:violation_mmfn}}\hfil
\subfloat[Autoware~\cite{kato2018autoware} collides with the front vehicle]{\includegraphics[height=0.16\textwidth,width=.22\textwidth]{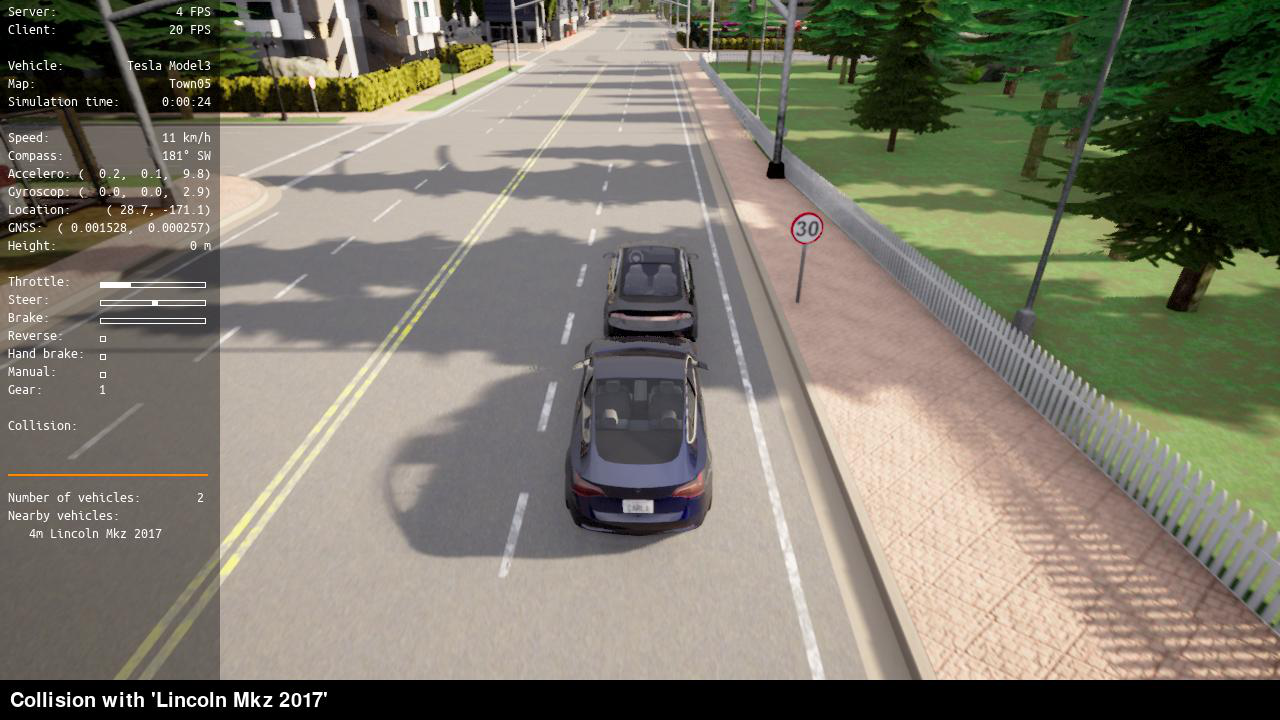}\label{fig:violation_autoware}}\hfil
\subfloat[LAV~\cite{chen2022lav} does not move even if the pedestrian has crossed the road]{\includegraphics[height=0.16\textwidth,width=.22\textwidth]{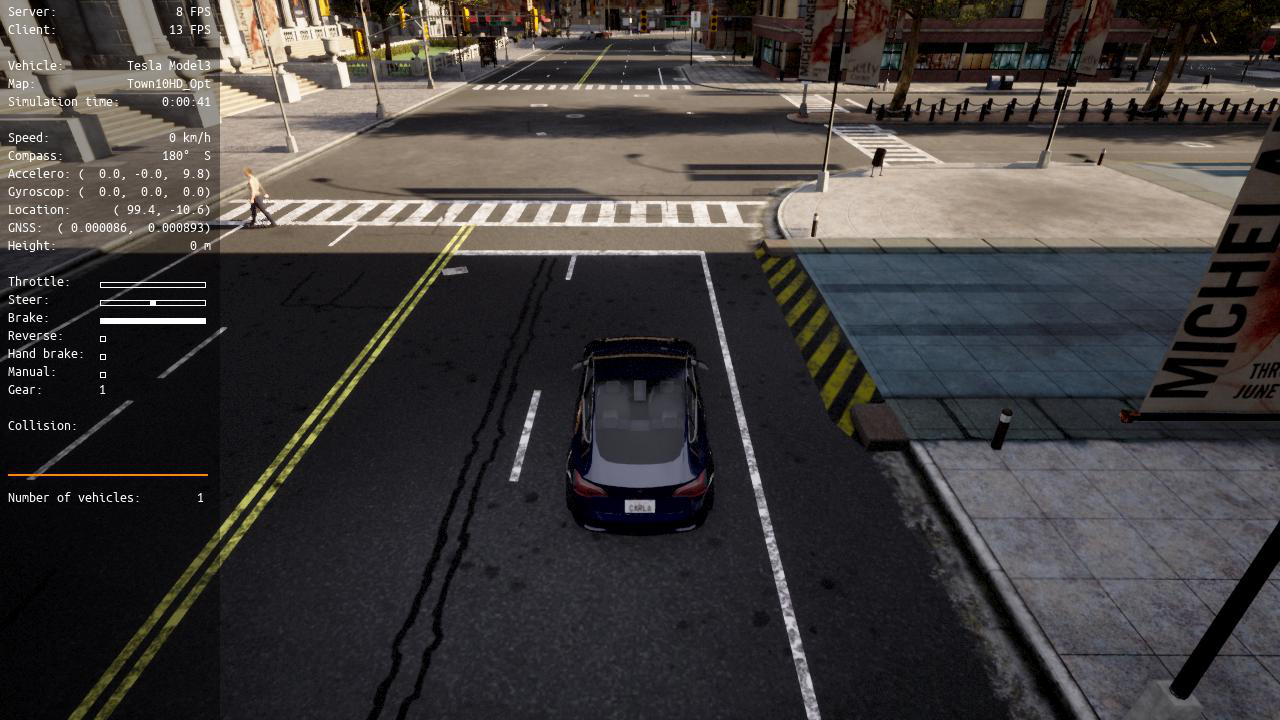}\label{fig:violation_person}}\hfil
\subfloat[Apollo system output: Fail to detect the pedestrian on the road]
{\includegraphics[height=0.16\textwidth,width=.22\textwidth]{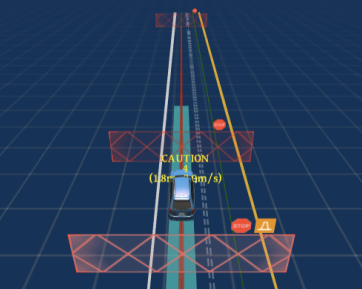}\label{fig:apollo_output}}\hfil

\caption{Examples of detected problems on ADSs in Carla and LGSVL}\label{figure}

\label{fig:all scenarios}
\end{figure*}

\begin{figure}
\subfloat[IDM~\cite{christiano2016transfer} collides with the other vehicle in roundabout]{\includegraphics[height=0.16\textwidth,width=.22\textwidth]{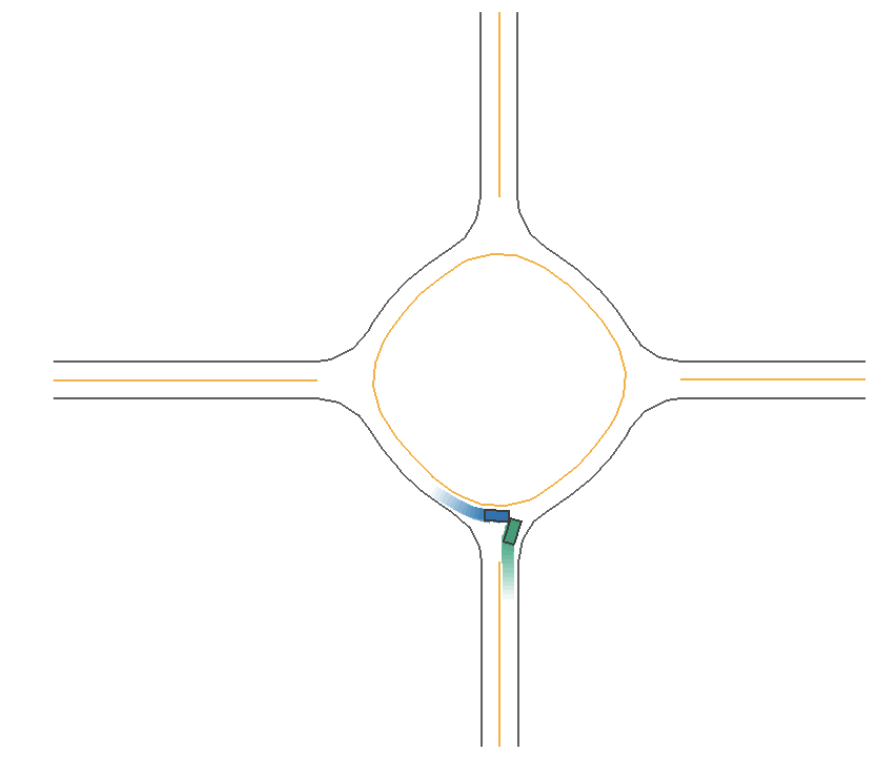}\label{fig:violation_idm}}\hfil
\subfloat[PPO~\cite{schulman2017proximal} collides with the opposite vehicle when turning left]
{\includegraphics[height=0.16\textwidth,width=.22\textwidth]{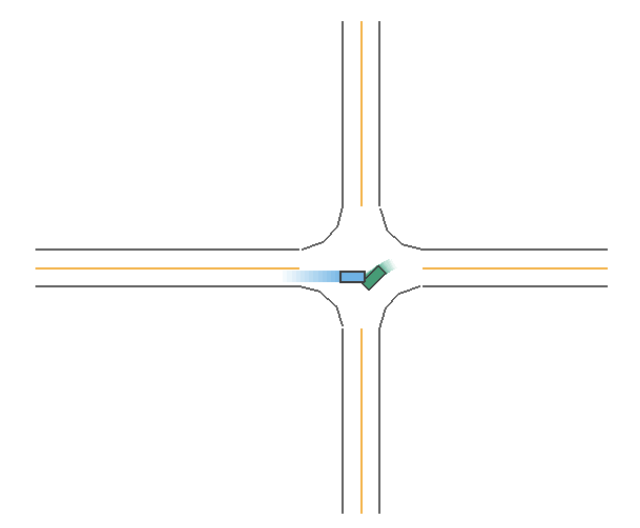}\label{fig:violation_ppo}}\hfil

\caption{Examples of detected problems on ADSs in MetaDrive}\label{figure}
\end{figure}


\begin{table}[]
\centering
\scalebox{1}{
\begin{tabular}{ccccc}
\hline
\rowcolor[rgb]{0.804,0.804,0.804} \textbf{Model} & \textbf{Rule violation} & \textbf{Collision} & \textbf{Timeout} \\
Auto~\cite{carla_auto}           & 75                                                                                     & 10                                    & {38}                                                                                                                      \\ 
\rowcolor[rgb]{0.898,0.898,0.898} MMFN~\cite{zhang2022mmfn}            & {72}                                                                                     & {43}                                    & {17}                                                                                                                       \\ 
LAV~\cite{chen2022lav}           & {46}                                                                                     & {2}                                    & {55}                                                                                                                        \\ 
\rowcolor[rgb]{0.898,0.898,0.898} Autoware~\cite{kato2018autoware}       & {68}                                                                                      & {88}                                     & {21}                                                                                                                     \\ 
Apollo 7.0~\cite{Apollo} & {7} & {8} & {4} \\
 \rowcolor[rgb]{0.898,0.898,0.898}IDM~\cite{christiano2016transfer} & {18} & {15} & {0} \\ 
PPO~\cite{schulman2017proximal} & {14} & {9} & {0} \\ \hline
\end{tabular}}
\caption{Detected violations for all ADSs under test}
\label{tab:violation_model}
\end{table}

From the human study in RQ2, some generated scenarios are voted as not matching with corresponding traffic rule descriptions. Such mismatch may cause the generated scenario not to correctly detect abnormal ADS behaviors. For example, some scenarios failed to accurately represent conditions like maintaining safe distances in rainy weather due to limitations in the simulator's rendering capabilities. To avoid such scenarios spoiling the experiment result, we first manually filtered out scenarios that did not match traffic rule descriptions. Then we used the left scenarios for detecting abnormal ADS behaviors. After manually filtering, 169, 88, and 27 scenarios generated in Carla, LGSVL, and Metadrive are valid for RQ3.

Table~\ref{tab:violation_model} shows the testing results of seven ADSs in the generated test scenarios. The generated test scenarios find many issues with these ADSs based on three metrics including rule violation, collision, and timeout. Among all ADSs, \textit{Apollo 7.0} performed best with the lowest rule violation and timeout problems. The main issue of \textit{Apollo 7.0} is the collision with slow-moved objects such as pedestrians, as shown in Figure~\ref{fig:violation_collision_apollo}.  \textit{MMFN}, \textit{Autoware} are prone to violate the traffic rules and lead to collisions, while the rule violation and collision are much fewer. \textit{Auto} and \textit{LAV} are less likely to cause collisions. However, \textit{LAV} leads to more timeout problems. To be more specific, the three ADSs \textit{Auto}, \textit{MMFN} and \textit{Autoware} have many violations of rules that require the ego vehicle to give way, as shown in Figures~\ref{fig:violation_roundabout} and ~\ref{fig:violation_intersection}; stop at the stop sign,  as shown in Figures~\ref{fig:violation_stop}; and increase following distances in different weather conditions. These rule violations mean that the three ADSs \textit{Autoware}, \textit{Auto}, and \textit{MMFN} cannot handle common driving scenarios to keep safe driving.  For two RL-based driving agents \textit{IDM} and \textit{PPO}, both of them tend to violate traffic rules and cause collisions. As shown in Figure~\ref{fig:violation_idm} and~\ref{fig:violation_ppo}, the two driving agents failed to give way and then caused collisions with other vehicles in the roundabout and in the intersection.





We further analyzed the test scenario recordings of the violation cases to figure out the underlying reasons. For \textit{Apollo}, we find it has difficulty detecting small and slow objects. For example, in Figure~\ref{fig:apollo_output}, the pedestrian was not detected by the perception system of \textit{Apollo 7.0}. This problem of Apollo 7.0 was also found in other research work~\cite{huai2023doppelganger}. For \textit{LAV}, we find the reason why it has few rule violations is that its behavior is prone to conservative. It often slows down or even stops when it reaches an intersection without traffic signals, even though there are no vehicles or pedestrians in the front, as shown in Figure~\ref{fig:violation_person}. This issue was reported to the authors and was confirmed.  With this driving strategy, \textit{LAV} reduces the risk of rule violations such as giving way or keeping a safe distance from the front vehicle and achieving low collision. However, such conservative behaviors lead to more timeout problems, since it takes a long time to give way or stop and then fails to reach the destination within the time limit. The reason for a low collision rate of \textit{Auto} is that it uses the ground-truth perception information rather than perceiving the environment by neural networks like other ADSs do. 


In addition, we found that \textit{MMFN}, and \textit{Autoware} cause more collisions. The ego vehicle controlled by \textit{MMFN} usually collides on the roadside, as shown in Figure~\ref{fig:violation_mmfn}. The problem is possibly in the localization module, which mistakenly thinks the ego vehicle is at an intersection. This problem has been confirmed by the authors of \textit{MMFN}. For \textit{Autoware}, the problem is possibly in the perception system, which has difficulty detecting static or slow-moving vehicles. Then the ego vehicle collides with the front vehicle, as shown in Figure~\ref{fig:violation_autoware}, which was similar to a reported issue in Github\footnote{Autoware issue: \url{https://github.com/carla-simulator/carla-autoware/issues/66}}.

\section{Related Work}
\label{sec:related_work}

\subsection{Scenario-based Test Generation}
AC3R (Automatic Crash Constructor from Crash Reports) \cite{gambi2019generating} constructs corner-case car crash scenarios for ADS testing using domain-specific ontology and the NLP technique dependency parsing~\cite{kubler2009dependency} to process police reports. However, these techniques often focus on keyword identification rather than text understanding~\cite{sha2003shallow}. In addition, AC3R requires well-formatted police reports as inputs. On the contrary, our approach, guided by our DSL (Section~\ref{sec:dsl}), and LLM-powered rule-parsing piepline, can understand comprehensive traffic rules and generate scenario representations, without needing rigid XML formatting as AC3R does. LawBreaker~\cite{sun2022lawbreaker} is a fuzzing testing framework that assesses whether ADSs breach traffic laws. It manually translates real-world traffic rules into DSLs and employs a fuzzing engine to generate test scenarios. Our work complements LawBreaker by presenting an automated pipeline to parse traffic rules to executable initial driving scenarios in simulation. 

Furthermore, search-based techniques \cite{gambi2019asfault, abdessalem2018testing, gambi2019automatically, luo2021targeting, zhong2022neural, tian2022mosat, calo2020generating, lu2022learning} examine road object interactions, like pedestrians and vehicles. They use optimization to pinpoint simulation corner cases for ADS testing. They employ fitness functions, such as minimum distance from the ego vehicle to other objects or time-to-collision, to guide the search of testing scenarios from seeding scenarios to safety-violation scenarios from generation to generation. For example, in~\cite{tian2022mosat}, a multi-objective fitness function is proposed to guide the search direction. The fitness function considers the minimal time-to-collision between the ego-vehicle and NPC vehicles, the maximal offset from the planned route to the driving trajectory of the ego-vehicle, and the maximal change of acceleration when the ego-vehicle is driving. Some works propose new search strategies to accelerate the creation of safety-violation cases. For example, in~\cite{zhong2022neural}, neural networks are used to predict and sort generated test cases that more likely caused violations during the search process. For these works, the common point is that the search usually begins with a random or DSL-described scenario. Our approach enhances these by auto-generating seeding scenarios consistent with traffic rules, potentially benefiting search-based testing or fuzzing. We plan to delve deeper into this in upcoming research.

Some existing works~\cite{guo2023scenario, montanari2021maneuver, weber2022toward, park2020creating} explore scenario reconstruction or replay based on real-world traffic data such as moving trajectories of vehicles. By analyzing such driving data, the corresponding scenarios are reconstructed in simulation platforms. Different from these works, our study focuses on generating test scenarios from traffic rules that do not provide real-world driving data of participants in driving scenarios.

\subsection{Scenario Description Language}
There exists substantial prior research focused on the generation of driving scenarios for testing ADSs.
As one kind of these methods, some DSLs are designed and proposed for ADS test generation. Fremont et al. \cite{Scenic} proposed Scenic, a DSL to describe scenarios (e.g., driving environment) using a probabilistic programming approach, to synthesize corner-case testing data.  Queiroz et al. \cite{geoscenario} proposed GeoScenario 
for the representation and evaluation of ADS test scenarios, by identifying relevant elements. However, it is not a trivial effort to design, understand, and master DSLs. AVUnit \cite{zhou2023specification} is a framework proposed for systematic ADS testing based on customizable correctness specifications. It employs two languages, SCENEST to specify dynamic scenarios, and AVSpec to formulate test oracles using signal temporal logic (STL). 
OpenScenario~\cite{openScenario}, which is the current scenario description standard based on XML, exhibits a rigid syntax and demands precise descriptions of numerous parameters (e.g., coordinates and orientations of objects). To implement a test scenario in such DSL, the tester must acquire additional domain knowledge and programming syntax.  Compared with existing DSLs, we propose a new DSL with 
simpler syntax, specifically tailored to describing test scenarios from traffic rules to facilitate automated test generation.

\section{Threats to Validity}
\label{sec:threat}

The \textbf{external validity} concerns the generalizability of our proposed test generation method.
In this paper, we tested five ADSs on two simulation platforms. MMFN~\cite{zhang2022mmfn} and LAV~\cite{chen2022lav}) were published at high-quality computer vision conferences such as CVPR and performed well in Carla Challenge~\cite{carla_challenge}. Auto~\cite{carla_auto} was developed based on the source code in the simulator Carla. Autoware~\cite{kato2018autoware} and Apollo are the largest open-source industry-level ADSs. IDM~\cite{christiano2016transfer} and PPO~\cite{schulman2017proximal} are advanced RL-based driving agents. For simulators, Carla, LGSVL and MetaDrive are widely used in ADS test generation works and they offer several distinct maps with high fidelity. Our experiment results show that the proposed method is generalizable to find violations of ADSs in different maps.

In this work, we applied the most relevant rules from the Texas driver handbook, because Texas is one of the most supportive states for the development of autonomous driving. Since 2017, the Department of Texas Transportation has published two laws about the operation of and regulation of autonomous vehicles~\cite{texasAVLaw}. Now there are about 15 autonomous driving companies such as Aruro and Waymo conducting their autonomous vehicle tests in Texas~\cite{texasAVCompany}. In future work, we plan to expand this study to test rules and regulations from various cities in the United States and other countries.

Using GPT-4 and our tailored rule-parsing pipeline, our rule parser accurately represented scenarios for 91.8\% of the benchmarked traffic rules, highlighting its capability to understand traffic rules and extract key information. In future work, we will incorporate traffic rules from other regions such as Asia. For scenario generation, we implemented a template scenario script and implemented it on two simulators Carla and LGSVL, which shows the generability of the proposed method on different simulators.

In terms of \textbf{internal validity}, the concern is that GPT-4 itself is powerful enough to generate scenario representation. Therefore, we conducted an ablation study to evaluate the effectiveness of the proposed rule-parsing pipeline. The experiment results show that the proposed pipeline achieves better performance than only using GPT-4.




The \textbf{construct validity} concerns the appropriateness of evaluation metrics used in the study. We conducted a human evaluation to assess whether generated testing scenarios match the descriptions of corresponding traffic rules. Human evaluation is widely used to measure the quality of test generation~\cite{deng2020rmt, gambi2019generating}. To measure agreement among human participants, we applied Fleiss' Kappa, a common metric for multiple human ratings. However, the original Fleiss' Kappa does not consider the weights of different voting categories. To address this limitation, we adapted Fleiss' Kappa by incorporating linear weights as described in~\cite{linearWeight}. In~\cite{landis1977measurement}, the authors suggested that Kappa scores between 0.61-0.80 indicate ``substantial agreement,'' while scores between 0.81-1.00 represent ``almost perfect agreement.'' It's important to note that these guidelines were originally proposed for two raters. In our study, with over 20 participants, achieving high agreement is more challenging. Given this context, our Kappa score of 0.68 on average demonstrates substantial agreement among a large number of participants.



\section{Discussion and Future Direction}
\subsection{Usefulness of Generated Scenarios by {\tool}}
Our paper proposes a novel framework, {\tool}, to automatically generate test scenarios for Autonomous Driving Systems (ADS) based on traffic rules. This framework addresses a critical need in the ADS industry: the efficient creation of diverse, representative, and challenging test scenarios. We rigorously tested {\tool} in multiple virtual environments, utilizing representative maps for rural, urban, and highway regions in Carla, LGSVL and MetaDrive simulators. These maps include diverse road networks such as intersections, roundabouts, and multi-lane roads, ensuring the generated scenarios are applicable across a broad spectrum of real-world driving conditions.

By deriving scenarios from established traffic regulations, {\tool} generates tests that represent fundamental and typical driving situations that both human drivers and ADS must navigate successfully, such as following, lane-changing, and giving way scenarios. Through generating these scenarios in different maps, {\tool} detected hundreds of abnormal behaviors in seven ADSs. Importantly, these detected violations can be mapped to real-world driving accidents. For instance, the National Highway Traffic Safety Administration identified the top five most frequent accident driving scenarios as Lead Vehicle Stopped, Vehicles Turning at Non-Signalized Junctions, Lead Vehicle Decelerating, Vehicles Changing Lanes - Same Direction, and Straight Crossing Paths at Non-Signalized Junctions~\cite{thorn2018framework}. Our work generated similar scenarios by following traffic rules related to maintaining safe distances and yielding right-of-way, which also resulted in ADS collisions with other actors. This alignment allows our framework to focus on high-risk scenarios, potentially uncovering critical flaws in ADS behavior that might lead to real-world accidents. 

In the future, we plan to extend {\tool} by incorporating multi-modal large language models to parse diverse sources such as accident reports containing text descriptions and sketches. This will enable direct reconstruction of accident scenarios to evaluate ADS performance. Additionally, {\tool} can be integrated with existing fuzzing-based test generation methods, providing meaningful initial seed scenarios that follow traffic rules or are based on accident reports. Fuzzing methods can then mutate these scenarios to discover more situations that may cause abnormal ADS behaviors. Recent work~\cite{dai2024sctrans} has demonstrated the effectiveness of the idea. We intend to pursue this direction in our future research.

\subsection{Trade-off between Manually and Automatically Generating Scenarios}
{\tool} can automatically generate test scenarios from traffic rules,  significantly streamlining the testing process for ADSs. This automation eliminates the need for extensive manual scenario creation, potentially saving considerable time and resources. The system's ability to rapidly produce a wide array of scenarios based on traffic rules ensures comprehensive coverage of various driving situations. However, the current limitations of LLMs require the design of a relatively simple DSL as middleware. This simplification means that the scenario script generator currently lacks the capability to define precise scenario parameters. For instance, testers cannot specify exact spawn points for vehicles on particular road segments. This limitation in granular control may restrict the ability to test highly specific scenarios or edge cases that require precise parameter settings.

Existing approaches to scenario creation often employ more complex DSLs, utilizing formats such as XML, temporal logic, or programming-like syntax. These methods allow testers to define scenarios with high precision, offering control over specific parameters. This level of detail is crucial for testing particular edge cases or replicating exact conditions observed in real-world incidents. However, the complexity of these DSLs presents a significant barrier to entry. For example, the OpenScenario user guide~\cite{openScenario} spans over 100 pages, requiring substantial time investment for testers to become proficient. Similarly, DSLs like LawBreaker~\cite{sun2022lawbreaker}, which use temporal logic, demand prerequisite knowledge in areas such as first-order logic. This complexity makes it challenging for less experienced testers to validate the correctness of their scenarios. An alternative approach involves directly writing test scenarios using simulator APIs. While this method offers precise control, it also requires extensive API knowledge and often involves a time-consuming trial-and-error process to determine specific parameters (e.g., the spawn location of a specific road segment). Our experience shows that manually creating a single test scenario using Carla API can take between 30 minutes and an hour, highlighting the significant time investment required for manual scenario creation.

Recognizing the strengths and limitations of both automated and manual approaches, we see potential in developing a hybrid solution. Our work can be extended to define a more powerful DSL capable of setting accurate parameters or incorporating existing DSLs. This evolution would aim to combine the efficiency of automation with the precision of manual control.

\subsection{Replicability of Applying LLMs}
The deployment of commercial LLMs for rule parsing incurs API-related expenses. However, these costs are manageable for both academic research and industry applications. Currently, the OpenAI API pricing~\footnote{OpenAI API Pricing: \url{https://openai.com/api/pricing/}} is approximately \$10 per million tokens. In our work, the average cost to parse a single rule is about \$0.05. We also evaluated the latest open-source LLM, Llama 3.1. In the experiments, we used the pretrained Llama3.1-8b and Llama3.1-70b models through Ollama~\cite{ollama} interface on an RTX4090 machine. Our experimental results show that Llama 3.1-70b (70 billion parameter version) achieves performance similar to GPT-3.5. This suggests that open-source LLMs could potentially match the performance of advanced commercial LLMs on ADS test generation tasks even without fine-tuning. With further training and advancements in prompt engineering techniques, open-source LLMs have the potential to achieve performance levels comparable to those of state-of-the-art commercial LLMs. To fully fine-tune an LLM model with 70 billion parameters, about 500GB of memory is needed. While using advanced LLM fine-tuning techniques such as LoRA~\cite{hu2021lora} and Q-LoRA~\cite{dettmers2024qlora}, 160GB and 48GB memories are needed respectively~\cite{llm_train}. Considering that a typical fine-tuning time is about three to five hours~\cite{llm_ft1, llm_ft2} and the rent price for a 48GB memory graphic card, A40 is about \$0.35 per hour, a fine-tuning experiment may cost \$1.05 to \$19.25, which is affordable for academic use. For deploying an LLM like Llama3.1-70b, 140GB memory is needed, which can be achieved by using 3 pieces of A40 GPU (about \$31,500) or 5 pieces of A100 GPU (about \$40,000). These costs are potentially manageable for ADS companies. Furthermore, for scenarios that do not require high-speed inference, LLM computation can be executed through hybrid GPU-normal memory modes or exclusively on standard memory, substantially reducing infrastructure expenses and enhancing accessibility for academic research.
We leave this as a promising direction for future research.

\section{Conclusion}
\label{sec:conclusion}
This paper presents {\tool}, a framework that auto-generates ADS test scenarios from natural language traffic rules. We introduce a DSL for these rules and an LLM-based parser to translate them into scenario representations. Using a hierarchical search algorithm, these are converted to scenario scripts. Experiments show TARGET effectively identifies rule violations and uncovers known issues in ADSs.
{\tool} represents the first attempt to leverage LLMs for automating the generation of driving scenarios directly from human rules. For future research directions, fuzzing and search-based algorithms can be developed to mutate scenario representations for generating more rule-violating test scenarios. The DSL for driving scenario representation can be expanded to support a more accurate scenario definition. Also, LLMs can be further studied, such as integrating multi-modality LLMs to improve the parsing accuracy of traffic rules.


\bibliographystyle{IEEEtran}
\bibliography{IEEEabrv,reference.bib}

\begin{IEEEbiography}[{\includegraphics[width=1in,height=1.25in,clip,keepaspectratio]{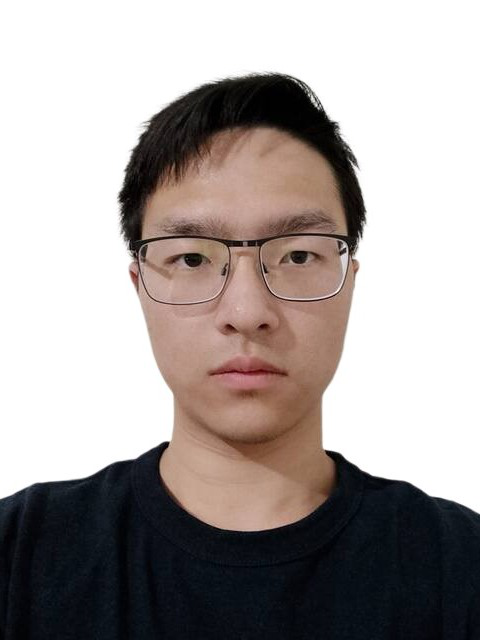}}]{Yao Deng} is a postdoc research fellow in Macquarie University. He obtained his Ph.D from Macquarie University in 2023, the Master of Research degree in Computing from Macquarie University in 2020, the Bachelor degree of Information Technology from Deakin University in 2018, and the Bachelor degree of Software Engineering from Southwest University in 2018. His current research interests include testing on autonomous driving systems and autonomous UAV systems.
\end{IEEEbiography}

\begin{IEEEbiography}[{\includegraphics[width=1in,height=1.25in,clip,keepaspectratio]{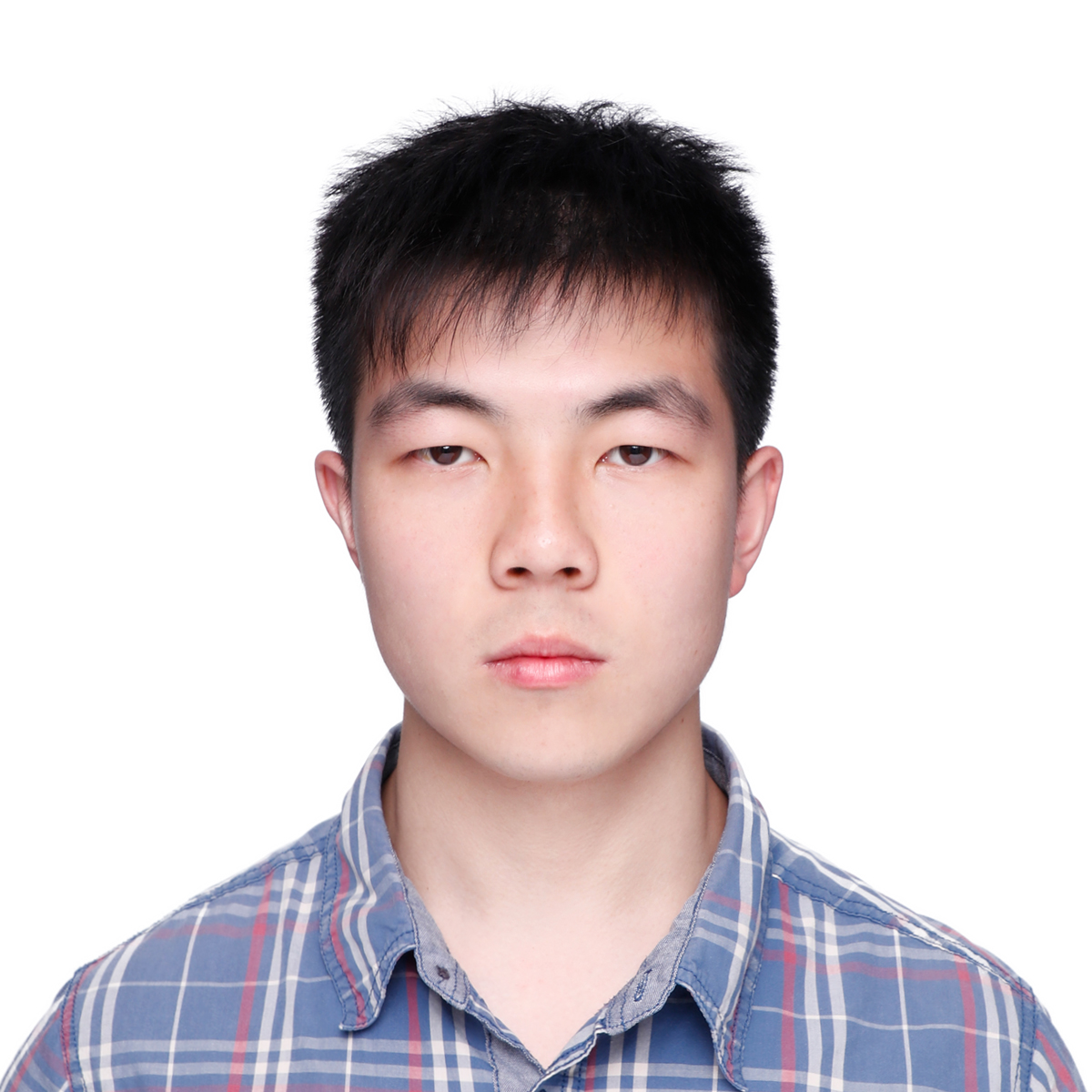}}]{Zhi Tu} received the B.E. degree in Computer Science and Technology from ShanghaiTech University, China, in June 2020, and the M.S. degree in Computer Science from the University of Southern California, USA, in August 2022. He is currently pursuing a Ph.D. degree in Computer Science at Purdue University. His research interests include computer vision, machine learning, and the testing of autonomous driving systems.
\end{IEEEbiography}

\begin{IEEEbiography}[{\includegraphics[width=1in,height=1.25in,clip,keepaspectratio]{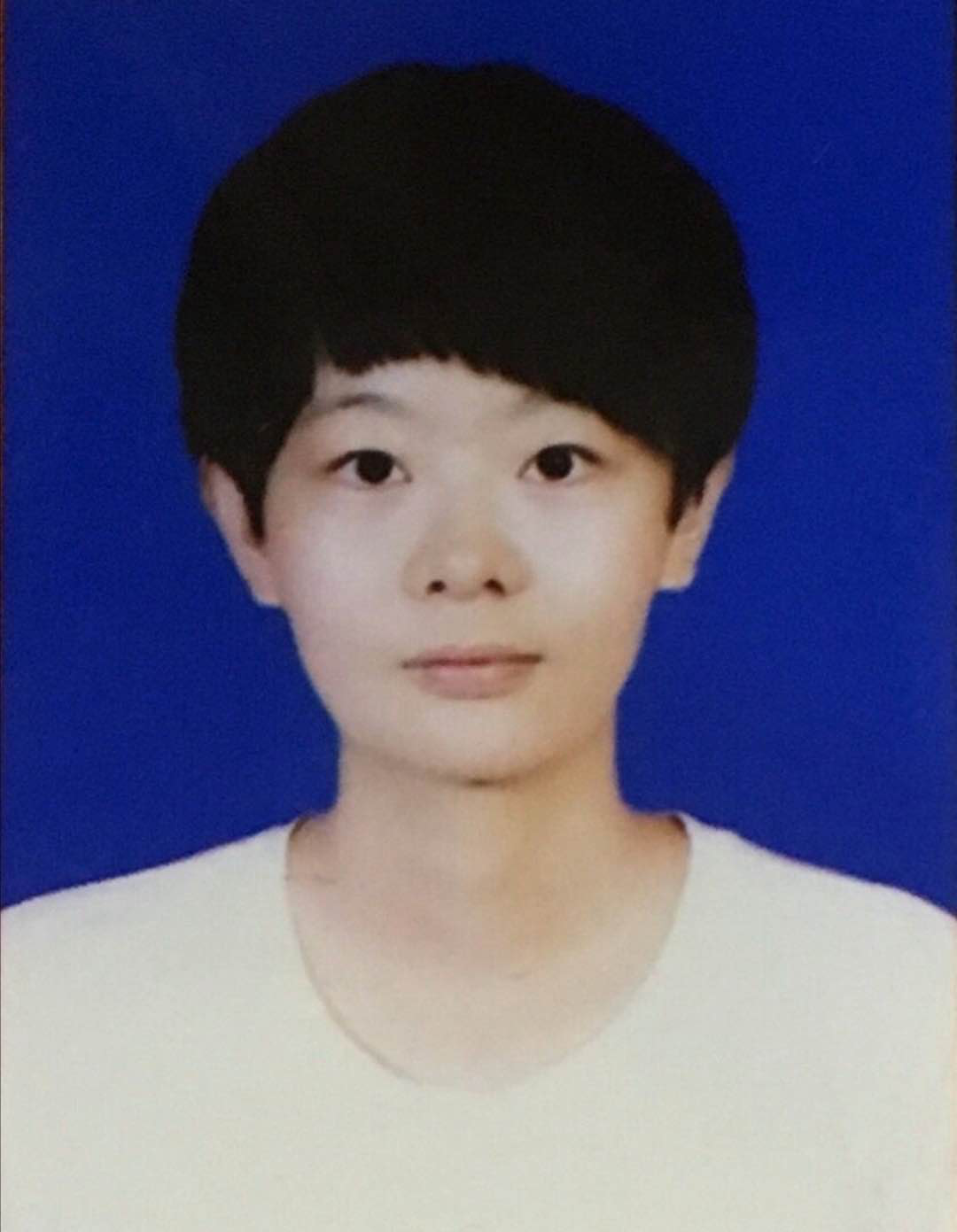}}]{Jiaohong Yao} is a Ph.D student at Macquarie University. She obtained her Master's degree in Information Technology from The University of Sydney in 2022, and Bachelor's degree in Applied Mathematics from China Jiliang University in 2020. Her current research topic is about autonomous drone landing and navigation.
\end{IEEEbiography}

\begin{IEEEbiography}[{\includegraphics[width=1in,height=1.25in,clip,keepaspectratio]{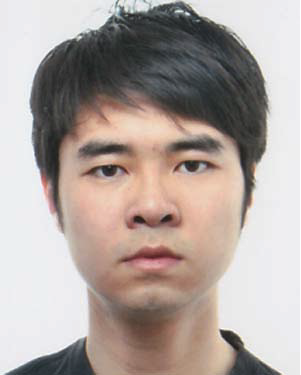}}]{Mengshi Zhang} received the BS degree in electronic engineering from Tsinghua University, in 2014 and the PhD degree from the Department of Electrical and Computer Engineering, University of Texas at Austin, in 2019. His research interests include fault localization, program repair, and machine-learningoriented software engineering.
\end{IEEEbiography}

\begin{IEEEbiography}[{\includegraphics[width=1in,height=1.25in,clip,keepaspectratio]{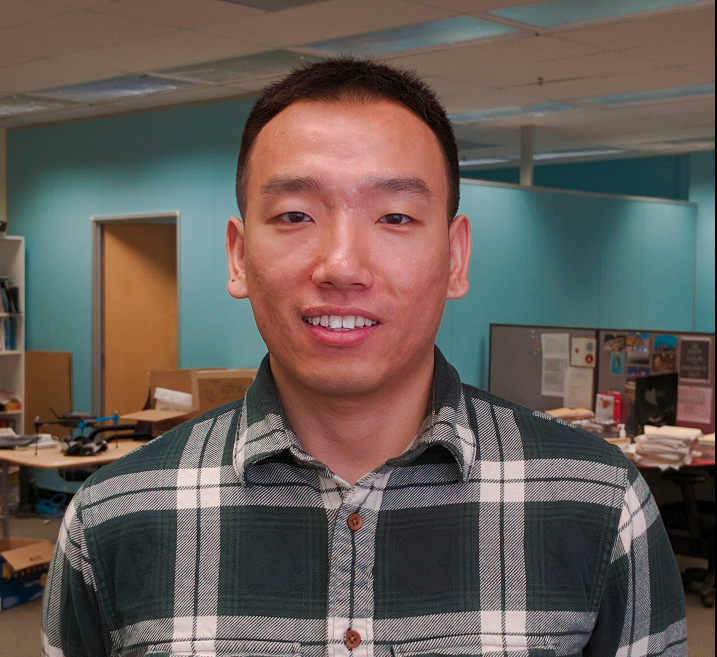}}]{Tianyi Zhang} is a Tenure-Track Assistant Professor in Computer Science at Purdue University. Prior to that, he was a Postdoctoral Fellow at Harvard University. He obtained his Ph.D. from University of California, Los Angeles in 2019 and his Bachelor's degree from Huazhong University of Science and Technology in 2013. His research interests include Software Engineering, Human-Computer Interaction, and Artificial Intelligence. In particular, his research focuses on building interactive systems that improve programming productivity and reduce coding barriers using AI-based technologies.
\end{IEEEbiography}

\begin{IEEEbiography}[{\includegraphics[width=1in,height=1.25in,clip,keepaspectratio]{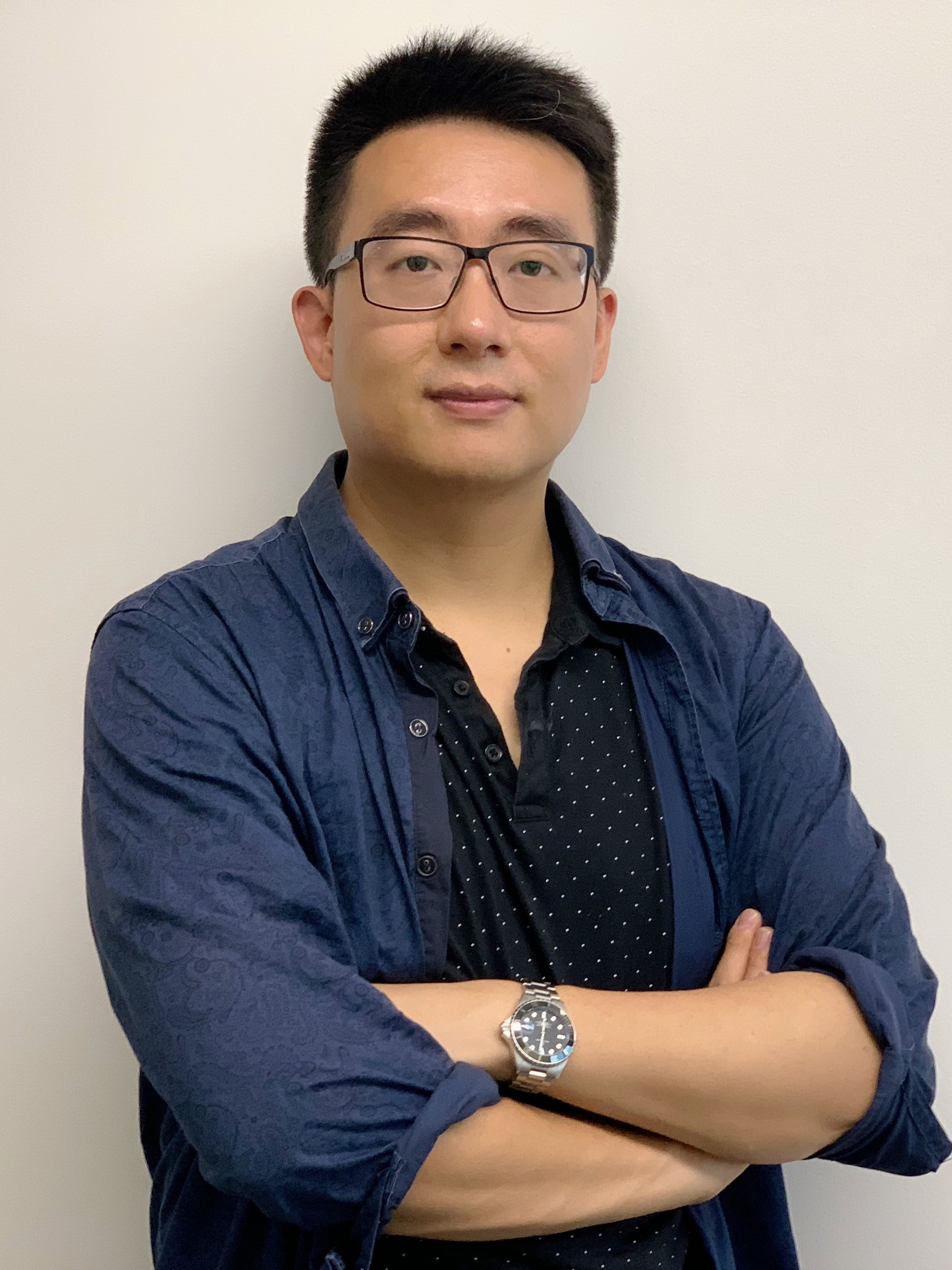}}]{Xi Zheng} earned his Ph.D. in Software Engineering from the University of Texas at Austin in 2015. Between 2005 and 2012, he was the Chief Solution Architect for Menulog Australia. Currently, he is an ARC Future Fellow and occupies several leadership roles at Macquarie University.His research areas include Cyber-Physical Systems Testing and Verification, Safety Analysis, Distributed Learning, Internet of Things, and the broader spectrum of Software Engineering.
\end{IEEEbiography}

\end{document}